\documentclass[aps,prc,twocolumn,groupedaddress,nofootinbib]{revtex4-2}

\usepackage{bm}
\usepackage{graphicx}
\usepackage{ascmac}
\usepackage{amsmath,amssymb}
\usepackage{braket}
\usepackage{url}
\usepackage{mathrsfs}
\usepackage{cancel}
\allowdisplaybreaks[1]

\begin{document}


\title{Eigenstates in coupled-channel scattering amplitude and their effects on spectrum}


\author{Takuma Nishibuchi}
\email[]{nishibuchi-takuma@ed.tmu.ac.jp}
\affiliation{Department of Physics, Tokyo Metropolitan University, Hachioji 192-0397, Japan}

\author{Tetsuo Hyodo}
\email[]{hyodo@tmu.ac.jp}
\affiliation{Department of Physics, Tokyo Metropolitan University, Hachioji 192-0397, Japan}


\date{\today}

\begin{abstract}
In general, discrete eigenstates such as resonances are represented by poles of the scattering amplitude, analytically continued to the complex energy plane. In multi-channel scattering, however, the Riemann surface becomes more complicated, leading to the emergence of various types of poles with distinct characteristics. In this study, we investigate the relationship between poles located on different Riemann sheets and analyze how they influence the observable spectra. In particular, we clarify the effect of the decay channel on the pole trajectory, where an $s$-wave bound state evolves into a resonance via a virtual state. It is shown that the quasibound state pole below the threshold does not continuously connected to the resonance pole above the threshold, and a kind of interchange of poles occurs. As a concrete example, we consider several models based on the chiral unitary approach that describe meson-baryon scattering amplitudes involving the $\Xi(1620)$ and $\Xi(1690)$ resonances. We examine their impact on the $\pi\Xi$ invariant mass distributions in the $\Xi_{c} \to \pi\pi\Xi$ decay, discussing how the pole structure manifests itself in experimental observables.
\end{abstract}


\maketitle


\section{Introduction}
\label{sec:intro}

Recent advances in experimental techniques have brought hadron spectroscopy into a new era~\cite{ParticleDataGroup:2024cfk}. Not only have exotic $XYZ$ states been discovered in systems containing heavy quarks such as $c$ and $b$~\cite{Hosaka:2016pey,Brambilla:2019esw}, but a number of exotic candidates have also been identified in the hadronic sector composed of $u$, $d$, and $s$ quarks~\cite{Guo:2017jvc,Hyodo:2020czb}. These hadrons are unstable via the strong interaction and are described as resonance states within the framework of scattering theory~\cite{Taylor,Rakityansky,Hyodo:2020czb,Mai:2022eur,Mai:2025wjb}. Resonances are represented as poles of the scattering amplitude analytically continued to the complex energy plane and can be treated on equal footing with bound states. Furthermore, depending on their position on the complex energy plane, the poles of the scattering amplitude are also referred to as virtual states and anti-resonance states.

On the other hand, actual exotic hadrons are generally coupled to multiple scattering channels. For example, the $\Xi(1620)$ and $\Xi(1690)$ resonances, which are the focus of this study, appear in coupled-channel meson-baryon scattering involving the $\pi\Xi$, $\bar{K}\Lambda$, $\bar{K}\Sigma$, and $\eta\Xi$ channels. In such coupled-channel problems, the analytic continuation of the scattering amplitude is defined on a multi-sheeted Riemann surface~\cite{Taylor}, leading to the appearance of poles corresponding to quasibound states, quasivirtual states, and their conjugate states~\cite{Nishibuchi:2023acl}. The mechanisms by which these poles are generated, and the ways in which they influence the spectra observed in experiments, are still not fully understood.

To investigate the origin of a given pole, it is useful to focus on the pole trajectory from the bound state to a resonance across the threshold. In the case of single-channel scattering, the pole trajectory under variation of the attractive interaction strength has been studied in Refs.~\cite{Hyodo:2014bda,Hanhart:2014ssa}. These studies have shown that when coupled to a partial wave with finite angular momentum, a bound state evolves directly into a resonance. In contrast, for $s$-wave scattering, the bound state does not become a resonance directly, but rather evolves via a virtual state in the intermediate step. By studying how such pole trajectories are modified when the system is coupled to open decay channels, and how a quasibound state evolves into a resonance, one can gain valuable insights into the origins of various poles in coupled-channel scattering.

Motivated by the above considerations, this study addresses the following two issues. First, from a theoretical perspective, we investigate the effect of the decay channels on the pole trajectory from an $s$-wave bound state into a resonance via a virtual state, using a simplified two-channel model. Second, we aim to clarify how the poles appearing on various Riemann sheets of the two-body scattering $T$-matrix affect physical observables, such as the actual scattering process and invariant mass distributions.

As a representative example of coupled-channel scattering, we focus on the meson-baryon scattering involving the $\Xi(1620)$ and $\Xi(1690)$ resonances. The strangeness $S = -2$ sector has recently seen significant accumulation of experimental data, including the $\pi^+\Xi^-$ invariant mass distribution in the $\Xi_c \to \pi\pi\Xi$ decay~\cite{Belle:2018lws}, as well as $K^{-}\Lambda$ correlation functions measured via femtoscopy techniques in high-energy Pb-Pb collisions~\cite{ALICE:2020wvi} and in $pp$ collisions~\cite{ALICE:2023wjz}. On the theoretical side, various scattering models based on the chiral unitary approach~\cite{Kaiser:1995eg,Oset:1997it,Oller:2000fj,Hyodo:2011ur,Hyodo:2020czb} have been proposed to describe this system~\cite{Ramos:2002xh,Garcia-Recio:2003ejq,Sekihara:2015qqa,Khemchandani:2016ftn,Feijoo:2023wua,Nishibuchi:2023acl,Li:2023olv,Sarti:2023wlg,Feijoo:2024qqg}.

This paper is organized as follows. In Sec.~\ref{sec:formulation}, we present the formulation of the scattering model employed in the subsequent analysis. Also, we provide explicit expressions for observables in terms of the scattering amplitude. In Sec.~\ref{sec:twochannel}, we investigate the impact of coupling to a decay channel on the poles and pole trajectories in a two-channel scattering system. In Sec.~\ref{sec:Xiresonances}, we apply this framework to the specific case of the $\Xi$ resonances, classify the poles appearing in several models proposed in previous studies, and examine their interrelations. Finally, we calculate the $\pi\Xi$ invariant mass distribution in the $\Xi_{c} \to \pi\pi\Xi$ decay and explore how the poles affect the observed spectrum in Sec.~\ref{sec:Xicecay}. 

\section{Formulation}
\label{sec:formulation}

In this section, we introduce the chiral unitary approach~\cite{Kaiser:1995eg,Oset:1997it,Oller:2000fj,Hyodo:2011ur,Hyodo:2020czb} to describe coupled-channel meson-baryon scattering amplitudes. The scattering equation for the amplitude $T_{ij}(W)$ at total energy $W$ is written in terms of the loop function $G_k(W)$ and the interaction kernel $V_{ij}(W)$ as
\begin{align}\label{eq:lseqchiral1}
T_{ij}(W) = V_{ij}(W) + \sum_k V_{ik}(W) G_k(W) T_{kj}(W),
\end{align}
where $i$, $j$, and $k$ are the channel indices. In the formulation based on the $N/D$ method~\cite{Oller:2000fj}, this scattering equation becomes an algebraic equation, so that the amplitude $T_{ij}(W)$ is fully determined by specifying the interaction kernel $V_{ij}(W)$ and the loop functions $G_k(W)$. Let $m_i$ and $M_i$ denote the meson and baryon masses in channel $i$, respectively. Physical scattering occurs at energies above the threshold, $W \geq m_i + M_i$. Furthermore, by analytically continuing $T_{ij}(W)$ to the complex energy plane, one can investigate resonance states, which appear as poles of the scattering amplitude.

In this study, we adopt the interaction kernel in the form
\begin{align} \label{eq:wtinteraction}
V_{ij}^{WT}(W) = -\frac{C_{ij}}{4f_i f_j}(2W - M_i - M_j)
\sqrt{\frac{M_i + E_i}{2M_i}} \sqrt{\frac{M_j + E_j}{2M_j}},
\end{align}
where $E_i$ is the energy of the baryon in channel $i$, and $f_i$ denotes the meson decay constant. By choosing appropriate values for the coupling coefficients $C_{ij}$, this expression reproduces the Weinberg-Tomozawa (WT) interaction, which corresponds to the leading-order term in chiral perturbation theory for meson-baryon scattering. Since $(2W-M_{i}-M_{j})$ is positive in the physical scattering region above the threshold and $V_{ij}$ corresponds to the Born approximation of the interaction potential in the nonrelativistic limit, $C > 0$ ($C < 0$) indicates an attractive (repulsive) interaction.

The loop function $G_k(W)$ is defined in terms of the four-momentum in the center-of-mass frame, $P^\mu = (W, \bm{0})$, as
\begin{align}\label{eq:loopfunc1}
G_k(W) = i \int \frac{d^4q}{(2\pi)^4} 
\frac{2M_k}{(P - q)^2 - M_k^2 + i0^+} 
\frac{1}{q^2 - m_k^2 + i0^+}.
\end{align}
The ultraviolet divergence in the $q$ integral is regularized using dimensional regularization, yielding the finite expression
\begin{align}
   &\quad G_k[W;a_k(\mu_{\mathrm{reg}})] \nonumber \\
   &=\frac{2M_k}{16\pi^2}\biggl[a_k(\mu_{\mathrm{reg}})+\ln\frac{m_kM_k}{\mu^2_{reg}}
   +\frac{M_k^2-m_k^2}{2W^2}\ln\frac{M_k^2}{m_k^2} \nonumber \\
   &\quad +\frac{\lambda^{1/2}(W^{2},M_{k}^{2},m_{k}^{2})}{2W^2}\nonumber \\
   &\quad \times\Bigl\{\ln(W^2-m_k^2+M_k^2+\lambda^{1/2}(W^{2},M_{k}^{2},m_{k}^{2})) \nonumber\\
   &\quad +\ln(W^2+m_k^2-M_k^2+\lambda^{1/2}(W^{2},M_{k}^{2},m_{k}^{2})) \nonumber \\
   &\quad -\ln(-W^2+m_k^2-M_k^2+\lambda^{1/2}(W^{2},M_{k}^{2},m_{k}^{2})) \nonumber \\
   &\quad -\ln(-W^2-m_k^2+M_k^2+\lambda^{1/2}(W^{2},M_{k}^{2},m_{k}^{2}))\Bigr\}\biggr], 
   \label{eq:loopfuncfinite}
\end{align}
where $a_k(\mu_{\mathrm{reg}})$ is the subtraction constant for channel $k$ at the regularization scale $\mu_{\rm reg}$. The K\"all\'en function is defined as
\begin{align}
\lambda(x,y,z)&=x^2+y^2+z^2-2xy-2yz-2zx
\label{eq:loopfuncfinitelambda}
\end{align}

Various observables can be calculated using the scattering amplitude described above. First, the nonrelativistic scattering amplitude $F_{ij}(W)$ is obtained from the $T$-matrix in Eq.~\eqref{eq:lseqchiral1} as
\begin{align}
F_{ij}(W) = -\frac{\sqrt{M_i M_j}}{4\pi W} T_{ij}(W). \label{eq:chiralf}
\end{align}
The scattering length $a_{0,i}$ in channel $i$ is then given by the diagonal component of $F_{ij}(W)$ at the threshold energy:
\begin{align}\label{eq:a0}
a_{0,i} = -F_{ii}(W = m_i + M_i).
\end{align}

Finally, following Refs.~\cite{Miyahara:2015cja,Miyahara:2016yyh,Oset:2016lyh}, we present the expression for the invariant mass distribution of the $MB$ system in the $\Xi_{c} \to \pi^{+} MB$ decay. Let $M_{\rm inv}$ be the invariant mass of the final meson-baryon channel $j$. Using the strangeness $S = -2$ meson-baryon scattering amplitude $T_{ij}$, the invariant mass distribution for channel $j$ is given by
\begin{align}
\frac{d\Gamma_j}{dM_{\rm{inv}}} = \frac{1}{(2\pi)^3} \frac{p_{\pi^+} \tilde{p}_j M_j}{M_{\Xi_c^+}} |\mathcal{M}_j|^2,
\label{eq:invdist}
\end{align}
where $p_{\pi^+}$ is the momentum of the emitted high-energy $\pi^+$, and $\tilde{p}_j$ is the meson momentum in the rest frame of the meson-baryon pair, given by
\begin{align}
p_{\pi^+} &= \frac{\lambda^{1/2}(M_{\Xi_c^+}^2, m_{\pi^+}^2, M_{\rm inv}^2)}{2M_{\Xi_c^+}}, \\
\tilde{p}_j &= \frac{\lambda^{1/2}(M_{\rm inv}^2, M_j^2, m_j^2)}{2M_{\rm inv}}.
\end{align}
The transition amplitude $\mathcal{M}_j$ is written as
\begin{align}
\mathcal{M}_j = V_P \left(h_j + \sum_i h_i G_i(M_{\rm inv}) T_{ij}(M_{\rm inv}) \right).
\label{eq:Minv}
\end{align}
Here, $V_P$ is a constant parameter representing the weak decay vertex, which is assumed to be approximately constant in the energy region of interest. The factors $h_i$, representing the weights of each meson-baryon channel in the final-state interaction, are determined by the wave function of the $\Xi_c$ and given as
\begin{align}
   h_{\pi^{0}\Xi^{0}} &= h_{\pi^{+}\Xi^{-}} = 0, \quad
   h_{\bar{K}^{0}\Lambda} = \frac{1}{\sqrt{6}}, \\
   h_{K^{-}\Sigma^{+}} &= 1, \quad 
   h_{\bar{K}^{0}\Sigma^{0}} = -\frac{1}{\sqrt{2}}, \quad
   h_{\eta\Xi^{0}} = -\frac{1}{\sqrt{3}}.
\end{align}

\section{Two-channel model}
\label{sec:twochannel}

\subsection{Model setup}
\label{subsec:setup}

Before applying the framework to actual hadronic systems, we investigate the response of poles to variations in the interaction using a simplified two-channel model, in order to understand the general properties of eigenstates near the threshold. Specifically, we study how the pole trajectory, which evolves from a bound state into a resonance via a virtual state, is affected by the coupling to a decay channel. 

For this purpose, we choose fictitious meson and baryon masses in each channel as
\begin{align}
   m_{1}&=150\ \text{MeV}, \quad M_{1}=1100\ \text{MeV}, \\
   m_{2}&=200\ \text{MeV}, \quad M_{2}=1300\ \text{MeV},
\end{align}
and define the interaction strengths using parameters $\alpha$ and $\beta$ as
\begin{align}
   C_{ij}=
   \begin{pmatrix}
   0 & \beta \\
   \beta & \alpha
   \end{pmatrix}.
\end{align}
with $f=110$ MeV. The subtraction constant in channel 1 is set as $a_{1}=-2.00$ at the regularization scale $\mu_{\rm reg}=630$ MeV. Here, $\alpha$ controls the interaction strength in channel 2, while $\beta$ determines the transition strength between the two channels. When $\beta = 0$, the channel coupling is switched off, and the system reduces to single-channel scattering in channel 2, with channel 1 undergoing independent free scattering. Therefore, we focus on the region near the threshold of channel 2 and examine the influence of the decay channel by varying the parameter $\beta$. 

In general, the analytic continuation of the two-channel scattering amplitude into the complex $W$ plane is defined on four Riemann sheets: [tt], [tb], [bt], and [bb], where t (b) denotes the first (second) Riemann sheet for each channel. Depending on which Riemann sheet a pole appears on, the corresponding eigenstate can be classified accordingly~\cite{Taylor,Nishibuchi:2023acl}.

\subsection{Bound state to resonance transition}
\label{subsec:BtoR}

First, we consider the single-channel problem near the threshold of channel 2 by setting $\beta=0$, which eliminates the channel coupling. In this case, because the choice of the Riemann sheet of channel 1 is irrelevant, a bound state in channel 2 appears at the same position on both the [tt] and [bt] Riemann sheets, while a virtual state and a resonance appear at the same position on the [tb] and [bb] sheets, respectively. By choosing an attractive interaction strength $\alpha = 4.0$ and a subtraction constant $a_2 = -3.00$, a single bound state appears as a pole at 1495 MeV on the [tt] and [bt] sheets, which we refer to as pole 1. Simultaneously, a virtual state emerges as a pole on the [tb] and [bb] sheets at 1436 MeV, which we refer to as pole 2 (see Table~\ref{tab:pole_beta0_a2}).

\begin{table}
   \caption{Pole positions in the two-channel model with $\alpha=4.0$ and $\beta=0$. The bound state ($B$) appears in the [tt] and [bt] sheets, while the virtual state ($V$), resonance ($R$), and anti-resonance ($\bar{R}$) appear in the [tb] and [bb] sheets. }
   \begin{ruledtabular}
   \begin{tabular}{llll}
   $a_{2}$ [dimensionless] & $-3.00$     & $-2.65$     & $-2.00$ \\
   \hline
   Pole 1 [MeV]            & 1495 ($B$) & 1494 ($V$) & $1509+ 80 i$ ($\bar{R}$) \\
   Pole 2 [MeV]            & 1436 ($V$) & 1474 ($V$) & $1509- 80 i$ ($R$)
   \end{tabular}
   \end{ruledtabular}
   \label{tab:pole_beta0_a2}
\end{table}

Next, we study the transition of the bound state pole into a resonance by varying the model parameters. One approach is to reduce the coupling constant $\alpha$, thereby weakening the attractive interaction. Another approach is to increase the subtraction constant $a_2$; it is known that for an attractive interaction with $C_{ii} > 0$, shifting the subtraction constant $a_i$ of channel $i$ in the negative direction enhances the attraction in that channel, while shifting it in the positive direction weakens it~\cite{Hyodo:2008xr,Ikeda:2011dx,Hyodo:2013iga}. In the following, we take the latter approach, while the results with the former approach is qualitatively same, as shown in Appendix~\ref{app:poletrajectory} for completeness.

As $a_{2}$ is increased from $-3.00$, pole 1 moves toward the threshold in the [tt] and [bt] sheets and becomes a virtual state in the [bt] and [bb] sheets at around $a_{2}\sim -2.80$. At the same time, pole 2 also moves toward the threshold in the [bt] and [bb] sheets. Further increasing $a_{2}$ causes pole 1 to move away from the threshold in the negative energy direction. Around $a_2 \sim -2.62$, it collides with pole 2. The collision point of these two poles is known as an exceptional point, where not only the pole positions but also their corresponding wave functions coalesce, unlike a simple degeneracy~\cite{PRE.61.929,Heiss:2012dx,Moiseyev}. After the collision, the poles move away from the real axis into the complex plane and eventually become a resonance and an anti-resonance at $a_2 = -2.00$. 

\begin{figure*}[tbp]
\centering
\includegraphics[width=8cm]{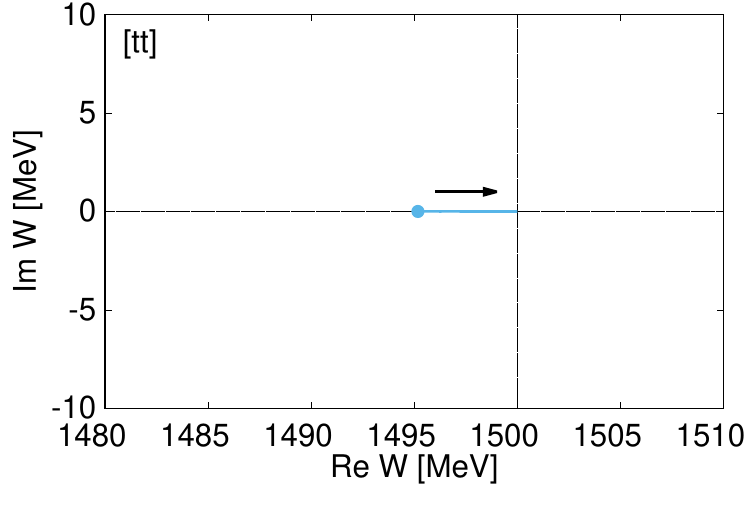}
\includegraphics[width=8cm]{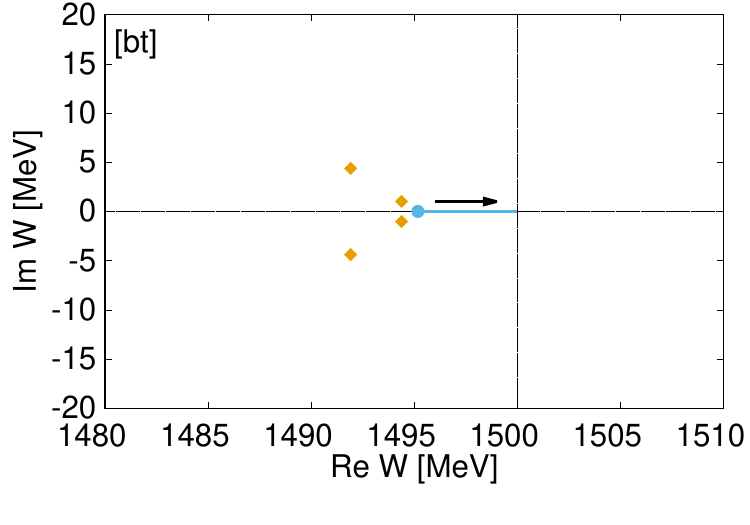}
\includegraphics[width=8cm]{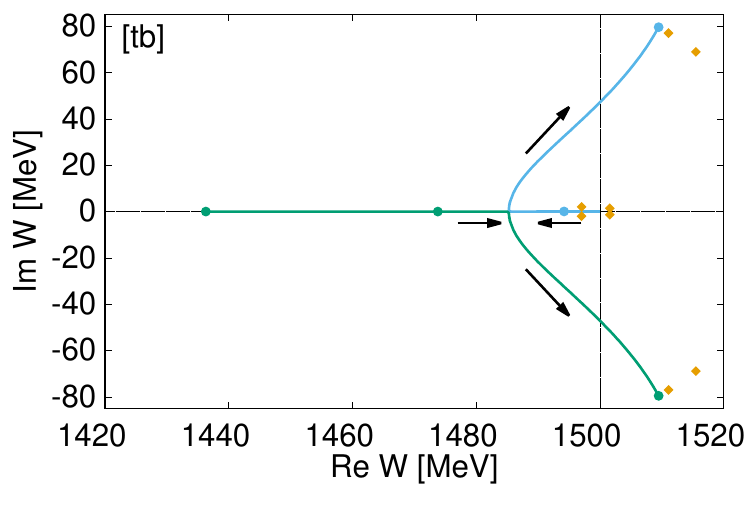}
\includegraphics[width=8cm]{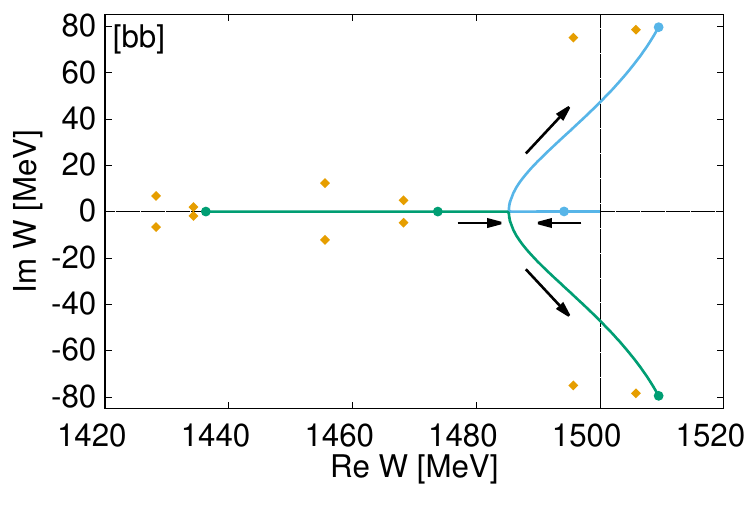}
\caption{Pole trajectories in the [tt] sheet (left top), [bt] sheet (right top), [tb] sheet (left bottom), [bb] sheet (right bottom), with the variation of the subtraction constant $a_{2}$ for $\alpha=4.0$ and $\beta=0$. The arrows indicate the direction of the pole movements as $a_{2}$ is increased. Poles at $a_{2}=-3.00$, $-2.65$, and $-2.00$ are marked by the circles. Squares represent the pole positions with $\beta=0.4$ and $\beta=0.8$ (finite channel coupling).}
\label{fig:pole_beta0_a2_E}
\end{figure*}

The trajectory of the poles with the variation of $a_{2}$ is shown in Fig.~\ref{fig:pole_beta0_a2_E} for each Riemann sheet. 
Pole positions at the representative values of the subtraction constant $a_{2}=-3.00$, $-2.65$, and $-2.00$ are summarized in Table~\ref{tab:pole_beta0_a2}. At these values, pole 1 (pole 2) is a bound state (virtual state), a virtual state (virtual state), and an anti-resonance (a resonance).\footnote{In the case with $\beta = 0$ where the poles pass through an exceptional point, it is not possible to distinguish which pole evolves into the resonance. Here, we tentatively assign pole 1 to the anti-resonance state, although the distinction between the two poles is not uniquely determined. We will show that this assignment is consistent with the results for $\beta\neq 0$.}

\begin{figure*}[tbp]
\centering
\includegraphics[width=8cm]{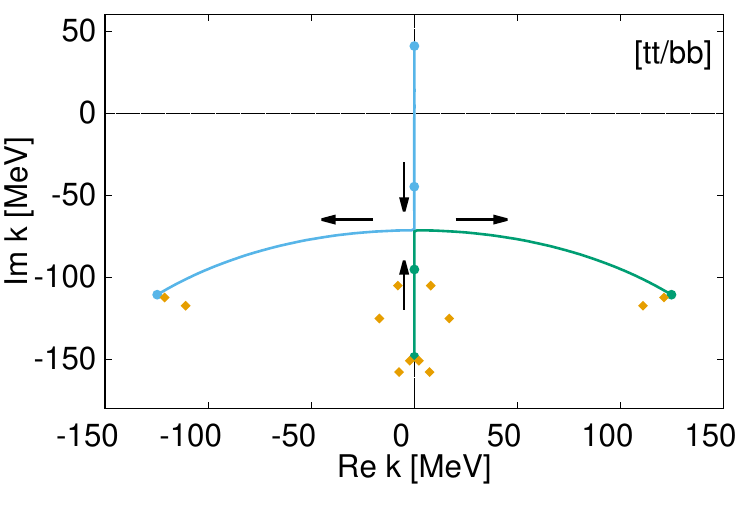}
\includegraphics[width=8cm]{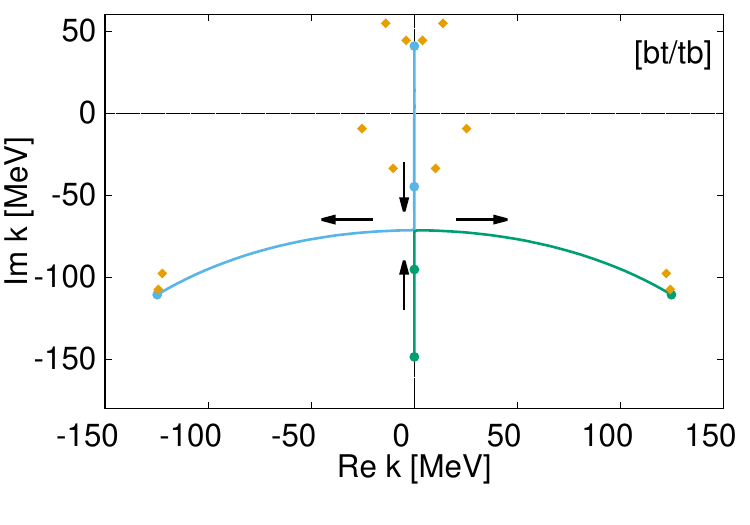}
\caption{Same with Fig.~\ref{fig:pole_beta0_a2_E} but in the complex momentum [tt/bb] plane (left) and [bt/tb] plane (right).}
\label{fig:pole_beta0_a2_k}
\end{figure*}

The pole trajectory can also be plotted on the complex momentum plane. The nonrelativistic momentum in channel 2 is defined as 
\begin{align}
   k=\sqrt{2\frac{M_{2}m_{2}}{M_{2}+m_{2}} (W-M_{2}-m_{2})} .
\end{align}
With the variable $k$, the Riemann sheets of channel 2 are distinguished by the upper and lower half planes, so it is sufficient to consider two combinations: one where the momentum of channel 1 is taken on the t-sheet (i.e., the [tt/tb] sheet), and the other on the b-sheet (i.e., the [tb/bb] sheet). However, as shown in Ref.~\cite{Nishibuchi:2023acl}, the [tt/tb] and [bt/bb] sheets contain a branch cut along the real $k$-axis, making them unsuitable for tracing the pole trajectory continuously. Therefore, we present the pole trajectories on the [tt/bb] and [tb/bt] sheets, where the lower half plane is exchanged, in Fig.~\ref{fig:pole_beta0_a2_k} left and right, respectively. In the case of $\beta=0$, the result does not depend on the choice of the Riemann sheet for channel 1, so the same pole trajectory appears on both sheets.

\subsection{Effect of channel coupling}
\label{subsec:decay}

Next, we investigate how the poles listed in Table~\ref{tab:pole_beta0_a2} are modified when a finite value of $\beta$ is introduced to include channel coupling. In the present model, the threshold energy of channel 1 is 1250~MeV, so all the poles in the table lie above this threshold. Therefore, we expect that the bound state becomes a quasibound state and the virtual state becomes a quasivirtual state once the channel coupling is introduced~\cite{Nishibuchi:2023acl}. In addition, when $\beta$ is finite, the symmetry between the [tt] and [bt] sheets, as well as between the [tb] and [bb] sheets, no longer holds, and thus the poles appear at different positions on all four Riemann sheets.

We begin by examining the bound state (pole 1) in the case of $\beta=0$ and $a_{2}=-3.00$, and study the effect of introducing a finite channel coupling. When $\beta$ is made finite, a pair of poles emerges symmetrically with respect to the real axis on the [bt] sheet. Among these, the pole with a negative imaginary part of the energy represents the quasibound state, which mainly contributes to physical scattering on the real energy axis.\footnote{Of course, this finite-width state appears as a resonance from the viewpoint of channel 1. However, in order to clarify its relation to the threshold energy of channel 2 and the distinction among different Riemann sheets, we follow the terminology of Ref.~\cite{Nishibuchi:2023acl} and refer to it as a quasibound state.} The finite decay width induced by channel coupling is reflected in the imaginary part of the pole position. The other pole, with a positive imaginary part, is referred to as an anti-quasibound state~\cite{Nishibuchi:2023acl}, as it is a conjugate pole of the quasibound state. On the other hand, the pole disappears in the [tt] sheet when $\beta$ is finite. In fact, no poles should appear in the [tt] sheet for finite $\beta$ since eigenstates with complex energy is forbidden on this sheet. The pole positions with $\beta=0.4$ and $\beta=0.8$ are shown in the right top panel of Fig.~\ref{fig:pole_beta0_a2_E}.

In this way, we find that switching on $\beta$ causes the pole on the [tt] sheet to disappear, while the single pole on the [bt] sheet splits into a pair of poles. In fact, we may consider that the two bound state poles on the [tt] and [bt] sheets for $\beta=0$ become the pair of poles on the [bt] sheet with finite $\beta$. When $\beta$ is finite, there is a unitary cut in the region above the channel 1 threshold $W > 1250~\mathrm{MeV}$, and the $W+i0^+$ region on the [tt] sheet is continuously connected to the $W-i0^+$ region on the [bt] sheet. This allows the bound state pole that was on the [tt] sheet for $\beta=0$ to move across the cut into the region $\mathrm{Im}\,W > 0$ on the [bt] sheet. The pole movement can also be observed in the complex momentum plane shown in Fig.~\ref{fig:pole_beta0_a2_k}. 

Next, we investigate the virtual states that appear on the [tb] and [bb] sheets at $\beta=0$. Similar to the bound state case, when a pole with a finite width exists on a given sheet, a conjugate pole must also appear at a position symmetric with respect to the real axis. Therefore, when $\beta$ is switched on, it is expected that a pair of poles will appear either on the [tb] sheet or the [bb] sheet, while disappearing from the other sheet. However, since both the [tb] and [bb] sheets can accommodate poles with finite widths, unlike the bound state case, this observation alone does not determine on which sheet the poles will appear.

By introducing a finite $\beta$ while focusing on the virtual state of pole 1 at 1494 MeV with $a_{2} = -2.65$, we find that a pair of poles appears on the [tb] sheet (see the left bottom panel of Fig.~\ref{fig:pole_beta0_a2_E}), and the corresponding pole disappears from the [bb] sheet. On the other hand, for the virtual state of pole 2 obtained at 1436 MeV with $a_{2} = -3.00$ and  at 1474 MeV with $a_{2} = -2.65$, a pair of poles emerges on the [bb] sheet when $\beta$ is finite (right bottom panel of Fig.~\ref{fig:pole_beta0_a2_E}). These results demonstrate that, in the case of a virtual state, the pair of poles can appear on either the [tb] or the [bb] sheet. As shown in the complex momentum plane in Fig.~\ref{fig:pole_beta0_a2_k}, it is natural from the continuity of the Riemann sheets that pole 1, which generates a pair of poles on the [bt] sheet when it is a bound state, produces a pair of poles on the [tb] sheet after it becomes a virtual state.

We now investigate the behavior of the resonance pole obtained at $a_{2} = -2.00$ under variation of $\beta$. As shown in Fig.~\ref{fig:pole_beta0_a2_E}, the pole lies at an identical position on the [tb] and [bb] sheets when $\beta = 0$. However, when $\beta$ is gradually introduced, the two poles follow different trajectories and become distinct. Among these, the pole on the [bb] sheet predominantly contributes to the physical scattering on the real energy axis, while the pole on the [tb] sheet is referred to as a shadow pole~\cite{Eden:1964zz}. In contrast to the case of bound or virtual states on the real axis, no branch cut appears at the resonance pole positions when $\beta$ is introduced, and thus the poles stay in the same Riemann sheet as $\beta=0$. Instead, the resonance pole and the shadow pole emerge separately. As seen in Fig.~\ref{fig:pole_beta0_a2_E}, the anti-resonance behaves symmetrically to the resonance. Again, in the momentum plane in shown Fig.~\ref{fig:pole_beta0_a2_k}, the movement of the poles can be understood by the continuity from the virtual state poles at $a_{2}=-2.65$; the pole on the [bt/tb] sheet moves upward, while that on the [tt/bb] sheet goes downward.

The effects of channel coupling on the pole structure discussed above can be summarized as follows:
\begin{itemize}
\item The bound state poles located on the real axis of the [tt] and [bt] sheets become a pair of quasibound and anti-quasibound states on the [bt] sheet.
\item The virtual state poles located on the real axis of the [tb] and [bb] sheets become a pair of quasivirtual and anti-quasivirtual states on either the [tb] or [bb] sheet.
\item The resonance poles located in the complex plane on the [tb] and [bb] sheets move differently and become resonance and shadow poles on the respective sheets.
\end{itemize}
In all cases, it should be noted that the total number of poles is conserved as $\beta$ is varied from 0 to finite values. In fact, by applying the argument principle to the phase of the scattering amplitude, it is shown that the number of poles remain unchanged under continuous deformation of model parameters, unless the pole encounter with a zero of the amplitude~\cite{Kamiya:2017pcq}.

\subsection{Pole trajectory with decay channel}
\label{subsec:trajectory}

\begin{figure*}[tbp]
\centering
\includegraphics[width=8cm]{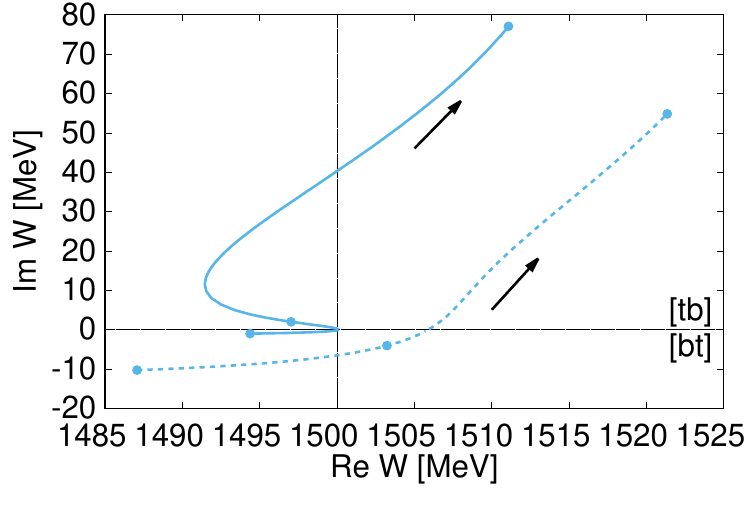}
\includegraphics[width=8cm]{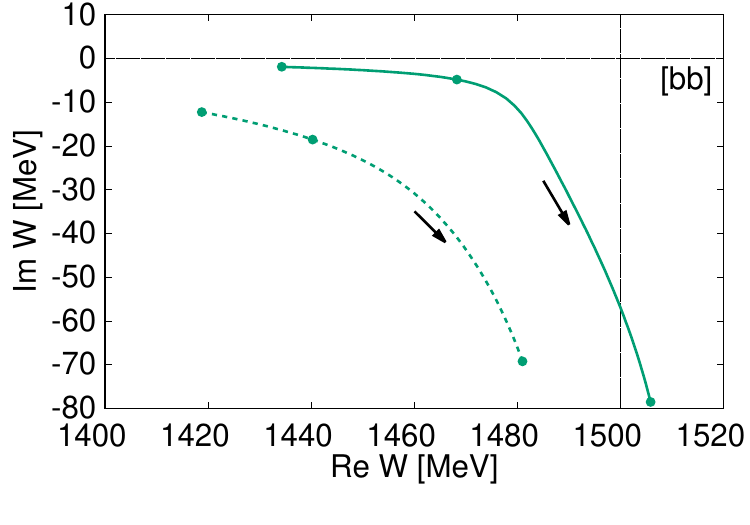}
\caption{Pole trajectories in the complex energy plane with the variation of the subtraction constant $a_{2}$ with $\beta=0.4$ (solid line) and $\beta=1.2$ (dashed line) for $\alpha=3.0$. The arrows indicate the direction of the pole movements as $a_{2}$ is increased. The left panel shows pole 1 on the [tb] sheet in the upper half plane and [bt] sheet in the lower half plane, while the right panel shows pole 2 on the [bb] sheet. The trajectories of the conjugate poles are omitted for simplicity. Poles at $a_{2}=-3.00$, $-2.65$, and $-2.00$ are marked by the circles.}
\label{fig:pole_beta_a2_E}
\end{figure*}

Finally, we show the trajectories of the poles for fixed finite values of the transition coupling $\beta$ with the variation of the subtraction constant $a_{2}$. In Fig.~\ref{fig:pole_beta_a2_E}, we show the results with $\beta=0.4$ and $\beta=1.2$ where the trajectories of the conjugate poles are omitted for simplicity. 

Pole 1 on the [tb] sheet at $a_{2}=-3.0$ (quasibound state) moves toward the threshold and, after crossing the cuts of both channels above the threshold energy of channel 2, moves into the [bt] sheet. Eventually, it becomes a shadow pole in the [bt] sheet at $a_{2}=-2.0$. With $\beta=0.4$, the trajectory of pole 1 closely resembles the case of $\beta=0$, leading to a local dip in the real part of the energy near the threshold. However, as $\beta$ increases, this non-monotonic structure disappears, and the trajectory evolves into a monotonic increase in the real part, leading smoothly to the shadow pole as seen in the $\beta=1.2$ case.

On the other hand, the resonance pole on the [bb] sheet at $a_{2}=-2.0$ originates from pole 2, which was a quasivirtual state on the [bb] sheet at $a_{2}=-3.0$. Given that at $\beta=0$, pole 1 was a bound state and pole 2 was its conjugate virtual state, this result indicates that a smooth connection from the subthreshold quasibound state to the above-threshold resonance does not occur. Instead, a kind of interchange between pole 1 (bound state) and pole 2 (virtual state) takes place. 

\begin{figure}[tbp]
\centering
\includegraphics[width=8cm]{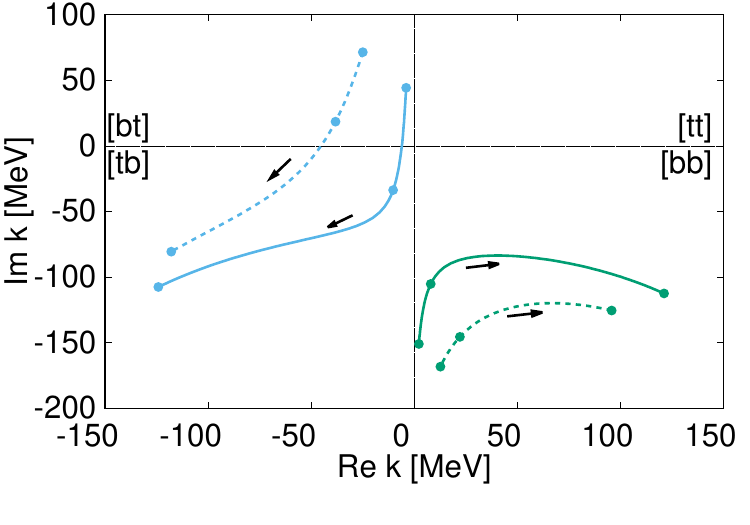}
\caption{Same with Fig.~\ref{fig:pole_beta_a2_E} but in the complex momentum plane. In this figure, the [tt/bb] plane is plotted in the Re $k>0$ region, whereas the [bt/tb] plane is plotted in the Re $k<0$ region.}
\label{fig:pole_beta_a2_k}
\end{figure}

By combining the left half of the complex momentum plane on the [bt/tb] sheet and the right half on the [tt/bb] sheet, we can simultaneously display the motion of pole 1 and pole 2, as shown in Fig.~\ref{fig:pole_beta_a2_k}. From this figure, it is clearly seen that pole 1, which was originally a quasibound state, evolves into an antiresonance, while pole 2, which started as a virtual state, moves closer to the threshold and eventually becomes the observable resonance. The behavior of poles around $a_{2}=-2.62$ can be understood as the level repulsion of the resonance states~\cite{PRE.61.929}, and is related to the exceptional point found at $\beta=0$.

\section{Application to $\Xi$ resonances}
\label{sec:Xiresonances}

\subsection{Poles in theoretical models}
\label{subsec:models}

In the following, we focus on the $\Xi(1620)$ and $\Xi(1690)$ resonances that appear in the $S = -2$ meson-baryon scattering. In the charge $Q=0$ sector, the coupled-channel scattering problem involves six channels: $\pi^0\Xi^0$, $\pi^+\Xi^-$, $\bar{K}^0\Lambda$, $K^-\Sigma^+$, $\bar{K}^0\Sigma^0$, and $\eta\Xi^0$. In the $Q=-1$ sector, the six channels are: $\pi^-\Xi^0$, $\pi^0\Xi^-$, $K^-\Lambda$, $K^-\Sigma^0$, $\bar{K}^0\Sigma^-$, and $\eta\Xi^0$. Accordingly, in the complex energy plane, one must specify either the t-sheet or the b-sheet for each of the six channels when discussing the Riemann surface structure, e.g., [bbtttt] sheet. Below, we analyze the $Q=0$ sector as an example. 

In this study, we employ the following models constructed based on the Weinberg-Tomozawa interaction: Set 1 from Ref.~\cite{Ramos:2002xh}, Set Fit from Ref.~\cite{Sekihara:2015qqa}, and Model 1 and Model 2 from Ref.~\cite{Nishibuchi:2023acl}. The corresponding interaction coefficients $C_{ij}$ for these models are summarized in Ref.~\cite{Nishibuchi:2023acl}. The hadron masses are taken from the values reported by the Particle Data Group~\cite{ParticleDataGroup:2024cfk}. The differences between the models originate primarily from the choice of the subtraction constants summarized in Table~\ref{tab:subtraction}, which, as discussed below, lead to different predictions for the $\Xi$ resonances. For the meson decay constants, we follow the conventions of each reference: Set 1, Model 1, and Model 2 use $f_\pi = f_K = f_\eta = 104.439$ MeV, while Set Fit adopts $f_\pi = 92.2$ MeV, $f_K = 1.2f_\pi$, and $f_\eta = 1.3f_\pi$.\footnote{Recently, more phenomenologically refined models have been developed, including those incorporating higher-order terms in chiral perturbation theory~\cite{Feijoo:2023wua,Feijoo:2024qqg}, fits to femtoscopy data~\cite{Sarti:2023wlg,Feijoo:2024qqg}, and fits to Belle data~\cite{Li:2023olv}. However, in this work, we are primarily interested in the qualitative origin of the poles, and therefore employ models based on the simpler Weinberg-Tomozawa interaction.}

\begin{table}
   \caption{Subtraction constant $a_{i}$ in theoretical models at the scale $\mu_{\rm reg}=630$ MeV. }
   \begin{ruledtabular}
   \begin{tabular}{lllll}
                                     & $a_{\pi\Xi}$ & $a_{\bar{K}\Lambda}$ & $a_{\bar{K}\Sigma}$ & $a_{\pi\Xi}$ \\
   \hline
   Set 1~\cite{Ramos:2002xh}         & $-2.00$      & $-2.00$              & $-2.00$             & $-2.00$ \\
   Set Fit~\cite{Sekihara:2015qqa}   & $-0.75$      & $-2.07$              & $-1.98$             & $-3.31$  \\
   Model 1~\cite{Nishibuchi:2023acl} & $-4.26$      & $-0.12$              & $-2.00$             & $-2.00$ \\
   Model 2~\cite{Nishibuchi:2023acl} & $-2.96$      & $\phantom{-}0.36$    & $-2.00$             & $-2.00$
   \end{tabular}
   \end{ruledtabular}
   \label{tab:subtraction}
\end{table}

For the charge $Q = 0$ sector, the positions of the resonance poles obtained in each model and their corresponding Riemann sheets are summarized in Table~\ref{tab:polephys}. In Set 1 in Ref.~\cite{Ramos:2002xh}, by taking the subtraction constants to be natural values $\sim -2$, the $\Xi(1620)$ is described as a broad resonance around the $\bar{K}\Lambda$ threshold (denoted as $z_{1}$ in Table~\ref{tab:polephys}). Set Fit in Ref.~\cite{Sekihara:2015qqa} focuses on the $\Xi(1690)$ and analyzes the $\bar{K}\Sigma$ and $\bar{K}\Lambda$ invariant mass distributions in the decay of $\Lambda_c$, resulting in the generation of a narrow-width pole near the $\bar{K}\Sigma$ threshold, which corresponds to the $\Xi(1690)$ resonance ($z_{3}$ in Table~\ref{tab:polephys}). Model 1 in Ref.~\cite{Nishibuchi:2023acl} reproduces the mass and decay width of $\Xi(1620)$ obtained in the analysis of $\Xi_c$ decays by the Belle experiment, producing a relatively narrow resonance located below the $\bar{K}\Lambda$ threshold ($z_{1}$ in Table~\ref{tab:polephys}). Model 2 in Ref.~\cite{Nishibuchi:2023acl} is constructed to reproduce the $K^{-}\Lambda$ scattering length measured by the ALICE experiment. In this case, the $\Xi(1620)$ is not described as a quasibound state but instead appears as a quasivirtual state above the threshold ($z_{2}$ in Table~\ref{tab:polephys}).

Since Set 1, Model 1, and Model 2 were constructed with a focus on the $\Xi(1620)$ near the $\bar{K}\Lambda$ threshold, a thorough search for poles in the $\Xi(1690)$ region has not been performed in these models. However, given that Set Fit yields a pole corresponding to the $\Xi(1690)$, it is possible that related poles also exist near the $\bar{K}\Sigma$ threshold in the other models. By exploring different Riemann sheets, we have found that all of Set 1, Model 1, and Model 2 contain a narrow-width pole on the \mbox{[tttbbt]} sheet ($z_{4}$ in Table~\ref{tab:polephys}), and a relatively broad-width pole on the [bbbbtt] sheet ($z_{5}$ in Table~\ref{tab:polephys}). Since all three models, Set 1, Model 1, and Model 2, commonly adopt $a_{\bar{K}\Sigma} = a_{\eta\Xi} = -2.00$, the scattering amplitudes exhibit similar behavior in the energy region near the $\bar{K}\Sigma$ threshold, leading to the appearance of poles at nearly identical positions.

\begin{table*}[tbp]  
\caption{Pole positions in theoretical models for charge $Q=0$ sector. The pole identified as describing either the $\Xi(1620)$ or $\Xi(1690)$ in each model is marked with an asterisk ($*$). }
\begin{ruledtabular}
\begin{tabular}{llll} 
Model & $z_1\ [{\rm{MeV}}]$ & $z_2\ [{\rm{MeV}}]$ & $z_3\ [{\rm{MeV}}]$ \\ \hline
Set 1~\cite{Ramos:2002xh} & $*1609\pm 140i$\ [bbtttt] &  $\phantom{*}1660\pm 164i$\ [bbtbbt] & $\phantom{*}1704\pm 41i$\ [tttbtt]\\
Set Fit~\cite{Sekihara:2015qqa} & $\phantom{*}1550\pm 281i$\ [bbtttt] &  $\phantom{*}1586\pm283i$\ [bbtbbt] &  $*1684\pm i$\ [bbbttt]\\ 
Model 1~\cite{Nishibuchi:2023acl} & $*1610\pm 30i$\ [bbtttt] &  $\phantom{*}1703\pm 22i$\ [bbtbbt] & $\phantom{*}1722\pm 24i$\ [tttbtt]\\
Model 2~\cite{Nishibuchi:2023acl} & $\phantom{*}1686\pm i$\ [bbtttt] & $*1726\pm 80i$\ [ttbttt] &  $\phantom{*}1705\pm 26i$\ [tttbtt] \\ \hline
Model &  $z_4\ [{\rm{MeV}}]$ & $z_5\ [{\rm{MeV}}]$ & \\ \hline
Set 1~\cite{Ramos:2002xh} & $\phantom{*}1680\pm 2i$\ [tttbbt] & $\phantom{*}1715\pm 63i$\ [bbbbtt] & \\ 
Set Fit~\cite{Sekihara:2015qqa} & $\phantom{*}1676\pm 6i$\ [bbbtbt] & $\phantom{*}1605\pm 13i$\ [bbbbtt] & \\
Model 1~\cite{Nishibuchi:2023acl} &  $\phantom{*}1684\pm 8i$\ [tttbbt] & $\phantom{*}1700\pm 91i$\ [bbbbtt] & \\
Model 2~\cite{Nishibuchi:2023acl} & $\phantom{*}1684\pm 3i$\ [tttbbt] & $\phantom{*}1687\pm 86i$\ [bbbbtt] & \\
\end{tabular}
\end{ruledtabular}
\label{tab:polephys}
\end{table*}%

\subsection{Interpolation of different models}
\label{subsec:interpolation}

As discussed above, different models yield different pole positions, but some of these poles are expected to have the same origin. For example, the $\Xi(1620)$ poles in Set 1 and Model 1 differ in their imaginary parts, but their real parts are close, suggesting that they may describe the same resonance. Moreover, Table~\ref{tab:subtraction} shows that the main differences between Model 1 and Model 2 lie in the values of $a_{\pi\Xi}$ and $a_{\bar{K}\Lambda}$, both of which are more negative in Model 1. This indicates a stronger attractive interaction in the $\pi\Xi$ and $\bar{K}\Lambda$ channels in Model 1. Therefore, there is a possibility that the quasibound state in Model 1 and the quasivirtual state in Model 2, both associated with the $\Xi(1620)$, are continuously connected, as illustrated in the two-channel analysis in Sec.~\ref{sec:twochannel}.

To examine the relation between the poles obtained in different models, we perform model interpolation. As discussed above, the differences between models arise from the subtraction constants $a_i$ and the meson decay constants $f_i$. Denoting the parameters of the initial model as \{$a_i^{I}$, $f_i^{I}$\} and those of the final model as \{$a_i^{F}$, $f_i^{F}$\}, we define the interpolation using a parameter $x$ as
\begin{align}
a_i(x) &= a_i^{I}(1 - x) + a_i^{F}x , \\
f_i(x) &= f_i^{I}(1 - x) + f_i^{F}x .
\end{align}
The model defined by \{$a_i(x)$, $f_i(x)$\} reproduces the results of the initial model at $x = 0$ and those of the final model at $x = 1$, allowing for a continuous connection between the two. In particular, by tracking the pole positions of the scattering amplitude as a function of $x$, we can study the trajectories of the poles between the two models. This continuous connection of poles obtained in different models provides insight into their origin.

\begin{figure}[tbp]
\centering
\includegraphics[width=8cm]{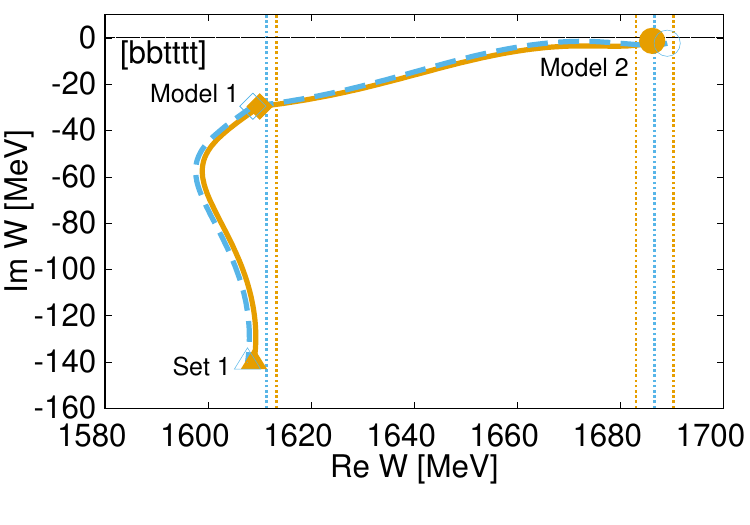}
\caption{Trajectory of pole $z_{1}$ during the interpolation between Set 1 (triangle), Model 1 (square), and Model 2 (circle). Interpolation with physical (isospin symmetric) model parameters is shown by the solid (dashed) line with the filled (open) symbols. Vertical dotted lines represent the threshold energies of isospin averaged $\bar{K}\Lambda$, $\bar{K}^{0}\Lambda$, $K^{-}\Sigma^{+}$, isospin averaged $\bar{K}\Sigma$, and $\bar{K}^{0}\Sigma^{0}$, from left to right. The trajectories of the conjugate poles are omitted for simplicity. }
\label{fig:z1}
\end{figure}

First, we perform an interpolation between Set 1, which produces a broad pole for the $\Xi(1620)$, and Model 1, which gives a narrow pole. Figure~\ref{fig:z1} shows the evolution of the pole $z_1$ by the solid line as the model parameters are continuously varied from Set 1 (triangle) to Model 1 (square). As seen in the figure, the broad pole $z_1$ of Set 1 moves closer to the real axis as the parameters change, and eventually becomes the narrow $z_1$ of Model 1. Since the pole does not cross any branch cut during the interpolation, both poles in these models lie on the same [bbtttt] Riemann sheet. Therefore, the poles for the $\Xi(1620)$ in these models share the same origin and can be interpreted as a $\bar{K}^0\Lambda$ quasibound state based on their Riemann sheet structure.

\begin{figure}[tbp]
\centering
\includegraphics[width=8cm]{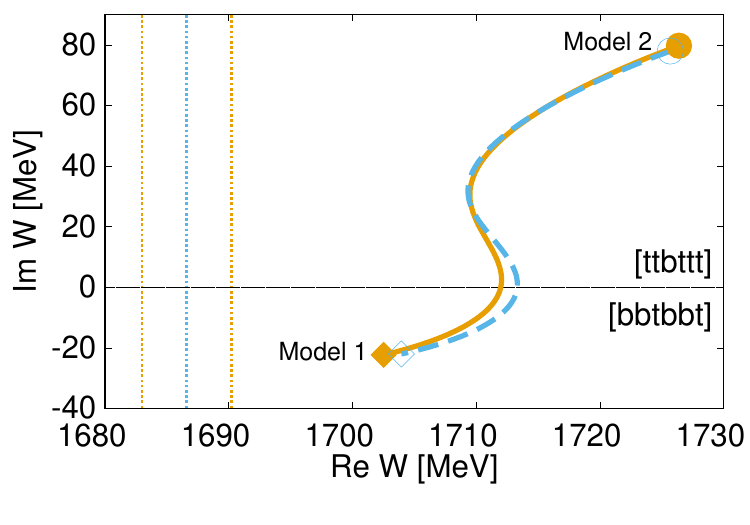}
\caption{
Trajectory of pole $z_{2}$ during the interpolation between Model 1 (square) and Model 2 (circle). Interpolation with physical (isospin symmetric) model parameters is shown by the solid (dashed) line. Vertical dotted lines represent the threshold energies of $K^{-}\Sigma^{+}$, isospin averaged $\bar{K}\Sigma$, and $\bar{K}^{0}\Sigma^{0}$, from left to right. The trajectories of the conjugate poles are omitted for simplicity. 
}
\label{fig:z2}
\end{figure}

As shown in Sec.~\ref{sec:twochannel}, a quasibound state located below the threshold on the [bt] sheet in the two-channel model evolves into a quasivirtual state on the [tb] sheet above the threshold when the attractive interaction is weakened. This suggests that the quasibound state $z_1$ in Model 1 in the [bbtttt] sheet may be continuously connected to the quasivirtual state $z_2$ in Model 2 in the [ttbttt] sheet. To investigate this possibility, we perform an interpolation between the pole $z_1$ of Model 1 and the pole $z_2$ of Model 2. The result is shown in Fig.~\ref{fig:z1} and Fig.~\ref{fig:z2} by the solid lines. As can be seen from Fig.~\ref{fig:z1}, the pole $z_1$ in Model 1, which is a quasibound state, moves continuously to higher energies across the $\bar{K}\Lambda$ threshold. However, it does not reach $z_2$ of Model 2 and instead becomes another pole in Model 2 located at $1684 - i$ MeV on the same [bbtttt] sheet. Although this pole is very close to the real axis, the [bbtttt] sheet in this energy region is not directly connected to physical scattering on the real axis, and thus the pole has little effect on the physical scattering amplitude. On the other hand, when interpolating continuously from $z_2$ of Model 2 on the [ttbttt] sheet to Model 1 (Fig.~\ref{fig:z2}), the pole crosses the real energy axis above the $\bar{K}\Sigma$ threshold, and moves to the [bbtbbt] sheet, where it remains.\footnote{When a pole crosses the real energy axis, it passes through the branch cuts of the open channels, and the t/b assignment for these channels is flipped. The trajectory of pole $z_2$ in Figure~\ref{fig:z2} crosses all cuts except for that of the highest-energy $\eta\Xi$ channel, resulting in t/b flips for the other five channels.} This indicates that the quasibound state in Model 1 and the quasivirtual state in Model 2 are not continuously connected and thus have different origins.

\begin{figure*}[tbp]
\centering
\includegraphics[width=8cm]{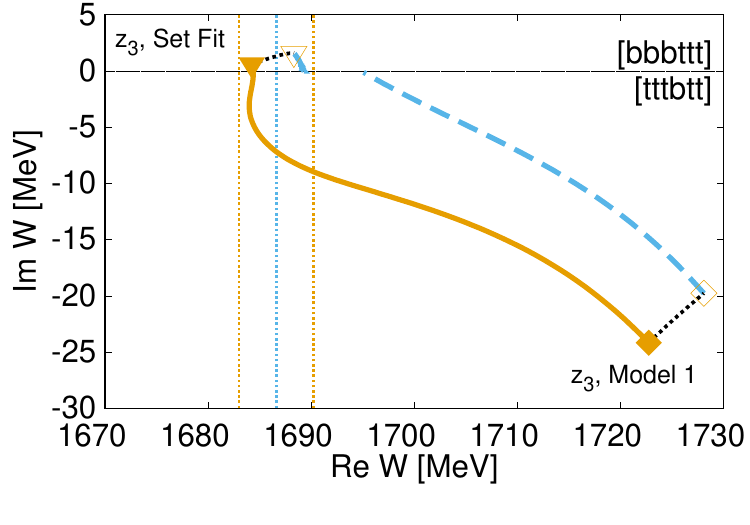}
\includegraphics[width=8cm]{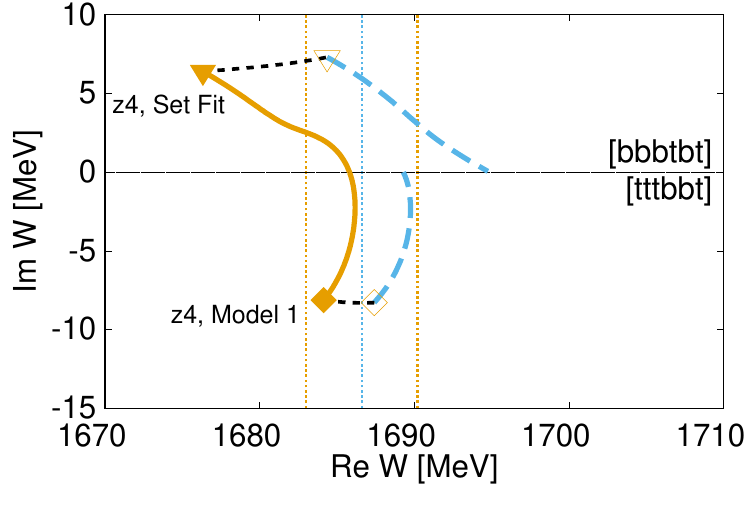}
\caption{
Trajectories of pole $z_{3}$ (left) and $z_{4}$ (right) during the interpolation between Set Fit and Model 1. Interpolation with physical (isospin symmetric) model parameters is shown by the solid (dashed) line. Vertical solid lines represent the threshold energies of $K^{-}\Sigma^{+}$, isospin averaged $\bar{K}\Sigma$, and $\bar{K}^{0}\Sigma^{0}$. The trajectories of the conjugate poles are omitted for simplicity. 
}
\label{fig:z3z4}
\end{figure*}

Similarly, we investigate the relationship between the $\Xi(1690)$ pole $z_3$ obtained in Set Fit and the poles $z_4$ and $z_5$ found in the other models. As mentioned above, the positions of $z_4$ and $z_5$ are almost identical in Set 1, Model 1, and Model 2. When interpolating between these models, it is confirmed that the poles $z_4$ and $z_5$ are indeed connected continuously with each other, as expected. However, when we interpolate from the $z_3$ pole found on the [bbbttt] sheet in Set Fit to Model 1, the pole crosses the real axis between the $K^-\Sigma^+$ and $\bar{K}^0\Sigma^0$ thresholds and moves to the [tttbtt] sheet. It remains at a location distinct from both $z_4$ and $z_5$, as shown by the solid line in the left panel of Fig.~\ref{fig:z3z4}. In the same way, the extrapolating $z_{4}$ in [tttbbt] sheet in Model 1 to Set Fit, the pole moves to the [bbbtbt] sheet by crossing the real axis between the $K^-\Sigma^+$ and $\bar{K}^0\Sigma^0$ thresholds (Fig.~\ref{fig:z3z4}, right). This indicates that, although $z_4$ and $z_5$ reside in the same energy region, they are states different from $z_3$.

In this way, we interpolate between different models and summarize the positions of the poles that are continuously connected in Table~\ref{tab:polephys}. Poles located on different Riemann sheets are those that have crossed the real axis during interpolation and moved to another sheet. Reflecting the values of the subtraction constants, we find that in the higher energy region, Set 1, Model 1, and Model 2 exhibit similar pole structures, while Set Fit shows distinct behavior. In contrast, in the lower energy region, Set 1, Model 1, and Model 2 differ from each other and provide different predictions.

\subsection{Effect of isospin symmetry breaking}
\label{subsec:isospin}

Among the poles listed in Table~\ref{tab:polephys}, $z_{3}$ of Model 1 is found on the [tttbtt] sheet, while $z_{4}$ of Set Fit appears on the [bbbtbt] sheet. These Riemann sheets differ in the treatment of the $K^{-}\Sigma^{+}$ and $\bar{K}^{0}\Sigma^{0}$ channels, which are isospin partners. In other words, these poles reside on Riemann sheets that arise due to the breaking of isospin symmetry. In this section, we discuss the effects of isospin symmetry breaking on the pole trajectories.

In the present model, the interactions and subtraction constants are constructed to be isospin symmetric, so that any isospin symmetry breaking arises solely from the hadron masses. To control the degree of isospin symmetry breaking, we introduce a parameter $y$ to describe the hadron masses with the physical masses $m_i^{\rm phys}$ and $M_i^{\rm phys}$ and the isospin-averaged masses $\bar{m}_i$ and $\bar{M}_i$ as
\begin{align}
m_i(y) &= m_i^{\rm phys}(1 - y) + \bar{m}_i y, \\
M_i(y) &= M_i^{\rm phys}(1 - y) + \bar{M}_i y.
\end{align}
At $y = 0$, the model uses the physical masses, while at $y = 1$, it becomes fully isospin symmetric. By continuously varying $y$ and tracing the pole positions, the effects of isospin symmetry breaking can be systematically investigated.

Starting from the pole positions listed in Table~\ref{tab:polephys}, we vary the parameter $y$ up to 1 to obtain the corresponding pole positions in the isospin symmetric case, summarized in Table~\ref{tab:poleiso}. For poles $z_{1}$ and $z_{2}$, the effect of isospin symmetry breaking is generally small, with shifts on the order of 1--2~MeV. In contrast, poles $z_{3}$, $z_{4}$, and $z_{5}$ show larger shifts ranging from a few MeV to about 10~MeV. This different behavior can be understood in terms of the distance from the thresholds that are split due to isospin symmetry breaking. In the present system, such threshold splittings occur in the $\pi\Xi$ channel (around 1450~MeV) and the $\bar{K}\Sigma$ channel (around 1690~MeV). Poles located closer to the $\bar{K}\Sigma$ threshold ($z_{3}$, $z_{4}$, $z_{5}$) are more significantly affected by isospin breaking.

The pole trajectories of $z_{1}$ and $z_{2}$ under the model interpolation in the isospin symmetric case are shown as dashed lines in Figs.~\ref{fig:z1} and \ref{fig:z2}, respectively. Since the isospin symmetry breaking had only a minor effect on the pole positions in each model, its impact on the pole trajectories is similarly small. As a result, the trajectories in the isospin symmetric case (dashed lines) closely follow those obtained using the physical masses (solid lines).

On the other hand, in the trajectories of $z_{3}$ and $z_{4}$ near the $\bar{K}\Sigma$ threshold, the isospin symmetry breaking leads to qualitatively different results. Figure~\ref{fig:z3z4} shows the pole positions of $z_{3}$ (left) and $z_{4}$ (right) for Set Fit and Model 1, respectively. The solid lines represent the model interpolations using the physical masses. The dotted lines show the trajectories with gradually imposing the isospin symmetry in hadron masses. Note that in the isospin symmetric case, the thresholds of $K^{-}\Sigma^{+}$ and $\bar{K}^{0}\Sigma^{0}$ become degenerate, and the region between these two thresholds disappears, where the poles $z_{3}$ and $z_{4}$ cross the real axis during the interpolation in the physical mass case. Therefore, the model interpolation with isospin-symmetric masses is not expected to reproduce the same trajectory as in the case with isospin symmetry breaking.

The result of the model interpolation under isospin symmetry is shown as dashed lines in Fig.~\ref{fig:z3z4}. When interpolating the $z_{3}$ pole of Set Fit to Model 1, the pole crosses the real axis above both the $K^{-}\Sigma^{+}$ and $\bar{K}^{0}\Sigma^{0}$ thresholds, moving across both cuts and transitioning from the [bbbttt] sheet to the [tttbbt] sheet, eventually reaching the $z_{4}$ pole of Model 1. Conversely, when interpolating the $z_{3}$ pole of Model 1, which lies on the [tttbtt] sheet, to Set Fit, it similarly crosses both cuts and moves to the [bbbtbt] sheet, finally arriving at the $z_{4}$ pole of Set Fit. This indicates that due to the breaking of isospin symmetry, $z_{3}$ and $z_{4}$ are poles that can transform into each other and can be interpreted as a kind of shadow poles.

\begin{table*}[tbp]  
\caption{Pole positions in isospin symmetric theoretical models. }
\begin{ruledtabular}
\begin{tabular}{llll} 
Model & $z_1\ [{\rm{MeV}}]$ & $z_2\ [{\rm{MeV}}]$ & $z_3\ [{\rm{MeV}}]$ \\ \hline
Set 1~\cite{Ramos:2002xh} & $1608\pm140i$\ [bbtttt] & $1659\pm 164i$\ [bbtbbt] & $1708\pm36i$\ [tttbtt] \\
Set Fit~\cite{Sekihara:2015qqa} & $1549\pm281i$\ [bbtttt]  &  $1585\pm283i$\ [bbtbbt] &  $1688\pm2i$\ [bbbttt] \\ 
Model 1~\cite{Nishibuchi:2023acl} & $1609\pm30i$\ [bbtttt] & $1704\pm22i$\ [bbtbbt] & $1728\pm20i$\ [tttbtt]\\
Model 2~\cite{Nishibuchi:2023acl} & $1689\pm 2i$\ [bbtttt] & $1726\pm 78i$\ [ttbttt] &  $1711\pm20i$\ [tttbtt] \\ \hline
Model &  $z_4\ [{\rm{MeV}}]$ & $z_5\ [{\rm{MeV}}]$ & \\ \hline
Set 1~\cite{Ramos:2002xh} & $1684\pm 3i$\ [tttbbt] & $1716\pm 58i$\ [bbbbtt] & \\ 
Set Fit~\cite{Sekihara:2015qqa} & $1684\pm7i$\ [bbbtbt] & $1603\pm12i$\ [bbbbtt] & \\
Model 1~\cite{Nishibuchi:2023acl} & $1687\pm8i$\ [tttbbt] & $1702\pm86i$\ [bbbbtt] \\
Model 2~\cite{Nishibuchi:2023acl} & $1687\pm 3i$\ [tttbbt] & $1687\pm79i$\ [bbbbtt] & \\
\end{tabular}
\end{ruledtabular}
\label{tab:poleiso}
\end{table*}%

\section{$\pi\Xi$ invariant mass distribution in $\Xi_{c}\to \pi\pi\Xi$ decay}
\label{sec:Xicecay}

\subsection{Comparison of mass distributions}
\label{subsec:comparison}

\begin{figure}[tbp]
\centering
\includegraphics[width=8cm]{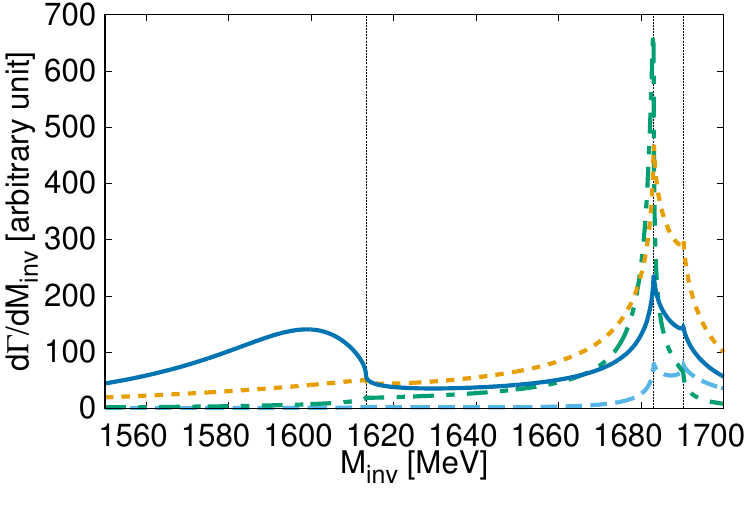}
\caption{
Invariant mass distribution of the $\pi^{+}\Xi^{-}$ system in the $\Xi_{c} \to \pi\pi\Xi$ decay. The 
dashed, dashed-dotted solid, and dotted lines represent the results from Set 1, Set Fit, Model 1, and Model 2, respectively. The vertical dotted lines indicate the thresholds of $\bar{K}^{0}\Lambda$, $K^{-}\Sigma^{+}$, and $\bar{K}^{0}\Sigma^{0}$ from left to right.}
\label{fig:Xicdecay}
\end{figure}

In this section, we investigate the impact of the scattering amplitude poles obtained in the previous section on observable quantities by calculating the $\pi^{+}\Xi^{-}$ invariant mass distribution in the $\Xi_{c} \to \pi\pi\Xi$ decay, which is experimentally accessible. By substituting the two-body scattering amplitudes $T_{ij}$ calculated in each theoretical model into the invariant mass distribution formula given in Eq.~\eqref{eq:invdist}, we quantitatively evaluate the influence of the poles listed in Table~\ref{tab:polephys} on the invariant mass spectrum. The results for Model 1 (solid line), Model 2 (dotted line), Set 1 (dashed line), and Set Fit (dash-dotted line) are shown in Fig.~\ref{fig:Xicdecay}.

We first focus on the $\Xi(1620)$ resonance near the $\bar{K}\Lambda$ threshold. Except for Model 2, the models have the pole $z_{1}$ located below the $\bar{K}\Lambda$ threshold. However, in Set 1 and Set Fit, where the imaginary part of the pole exceeds 100~MeV, no peak structure appears on the real axis. As a result, only Model 1 exhibits a prominent peak structure of $\Xi(1620)$ at energies below the $\bar{K}\Lambda$ threshold.

Next, we examine the $\Xi(1690)$ resonance near the $\bar{K}\Sigma$ threshold. In this energy region, the narrow-width poles are $z_{3}$ and $z_{4}$, but as shown in Table~\ref{tab:polephys}, none of these poles appear on the physically relevant Riemann sheet in any of the models. Here, the ``physically relevant Riemann sheet'' refers to the Riemann sheet directly connected to the real-energy scattering process~\cite{Nishibuchi:2023acl}. Specifically, for ${\rm Re}\ z < 1683\ \text{MeV}$ (below the $\bar{K}^{-}\Sigma^{+}$ threshold), the relevant sheet is [bbbttt]; for $1683\ \text{MeV} < {\rm Re}\ z < 1690\ \text{MeV}$ (between the $K^{-}\Sigma^{+}$ and $\bar{K}^{0}\Sigma^{0}$ thresholds), it is [bbbbtt]; and for ${\rm Re}\ z > 1690\ \text{MeV}$ (above the $\bar{K}^{0}\Sigma^{0}$ threshold), it is [bbbbbt]. The only pole that appears on the physically relevant Riemann sheet is $z_{5}$ in Model 2, but this pole has a large imaginary part of 86~MeV. As a result, none of the models exhibit a clear peak structure associated with the $\Xi(1690)$ resonance.

On the other hand, Fig.~\ref{fig:Xicdecay} shows a prominent cusp structure at the $\bar{K}\Sigma$ threshold. In Model 1 (solid line), a cusp of comparable height to the $\Xi(1620)$ resonance peak is observed at the $\bar{K}\Sigma$ threshold. Furthermore, the threshold cusp structures in Model 2 and Set Fit exhibit even larger amplitudes than that of Model 1. While the cusp structure in Set 1 is smaller in absolute magnitude compared to the other models, this can be attributed to the overall small magnitude of the invariant mass distribution in Set 1. In fact, the mass distribution of Set 1 near the $\bar{K}\Sigma$ threshold is relatively enhanced compared to other energy regions. Thus, a strong cusp structure at the $\bar{K}\Sigma$ threshold is confirmed in all models. The origin of this feature will be investigated in detail in the next section.

\subsection{Threshold cusp at $\bar{K}\Sigma$ threshold}
\label{subsec:cusp}

It is generally known that a strong cusp structure is enhanced when the absolute value of the scattering length in the threshold channel is large~\cite{Baru:2004xg,Sone:2024nfj}. The scattering lengths for the $K^{-}\Sigma^{+}$ and $\bar{K}^{0}\Sigma^{0}$ channels, evaluated using Eq.~\eqref{eq:a0}, are summarized in Table~\ref{tab:a0}. While the scattering length of the $K^{-}\Sigma^{+}$ channel is slightly larger than the typical hadronic scale ($\sim 1$~fm), it is not large enough to expect a pronounced cusp enhancement. The scattering length of the $\bar{K}^{0}\Sigma^{0}$ channel is comparable to the typical hadronic scale. Among the four models, the absolute values of the scattering lengths are smallest in Model~1, followed by Model~2 and Set~1, and largest in Set~Fit. This ordering does not correspond to the strength of the cusp structures observed in the invariant mass distributions shown in Fig.~\ref{fig:Xicdecay}. Therefore, it can be concluded that the enhancement of the cusp structure in the invariant mass distributions is not driven by the scattering lengths in the $\bar{K}\Sigma$ channels.

\begin{table*}[tbp]\label{tab:a0}
\caption{Scattering lengths for the $K^{-}\Sigma^{+}$ and $\bar{K}^{0}\Sigma^{0}$ channels.}
\begin{center}
\begin{ruledtabular}
\begin{tabular}{lllll}
Model & $a_{0,K^-\Sigma^+}\ [{\rm{fm}}]$ & $|a_{0,K^-\Sigma^+}|\ [{\rm{fm}}]$ & $a_{0,\bar{K}^0\Sigma^0}\ [{\rm{fm}}]$ & $|a_{0,\bar{K}^0\Sigma^0}|\ [{\rm{fm}}]$ \\ \hline
Set 1~\cite{Ramos:2002xh}         & $-1.09-1.19i$           & $1.61$ & $-0.42-0.68i$           & $0.80$ \\
Set Fit~\cite{Sekihara:2015qqa}   & $\phantom{-}0.70-1.60i$ & $1.75$ & $\phantom{-}0.54-0.82i$ & $0.98$ \\
Model 1~\cite{Nishibuchi:2023acl} & $-0.63-1.05i$           & $1.22$ & $-0.23-0.56i$           & $0.61$ \\
Model 2~\cite{Nishibuchi:2023acl} & $-0.70-1.39i$           & $1.56$ & $-0.23-0.75i$           & $0.78$ \\
\end{tabular}
\end{ruledtabular}
\end{center}
\label{default}
\end{table*}%

\begin{figure}[tbp]
\centering
\includegraphics[width=8cm]{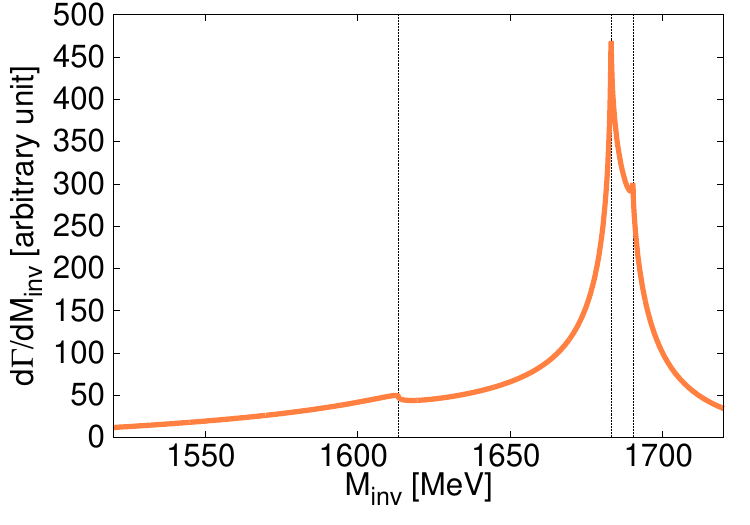}
\caption{
Invariant mass distribution of the $\pi^{+}\Xi^{-}$ system in the $\Xi_{c} \to \pi\pi\Xi$ decay with Model 2. The vertical dotted lines indicate the thresholds of $\bar{K}^{0}\Lambda$, $K^{-}\Sigma^{+}$, and $\bar{K}^{0}\Sigma^{0}$ from left to right.}
\label{fig:XicdecayM2}
\end{figure}

In the following, we examine the origin of the cusp structure using Model~2 as a representative example. The invariant mass distribution for Model~2 alone is shown in Fig.~\ref{fig:XicdecayM2}, where a prominent cusp structure at the $\bar{K}\Sigma$ threshold is clearly observed. The $M_{\rm inv}$ dependence of the mass distribution is mainly governed by the two-body scattering amplitudes $T_{ij}(M_{\rm inv})$ of Model~2. However, previous studies of the two-body scattering amplitudes have not reported a strong $\bar{K}\Sigma$ cusp. In fact, Fig.~10 of Ref.~\cite{Nishibuchi:2023acl} presents the elastic scattering amplitude for the $K^{-}\Lambda$ channel, in which the effect of the $\bar{K}\Sigma$ cusp is hardly visible.

\begin{figure}[tbp]
\centering
\includegraphics[width=8cm]{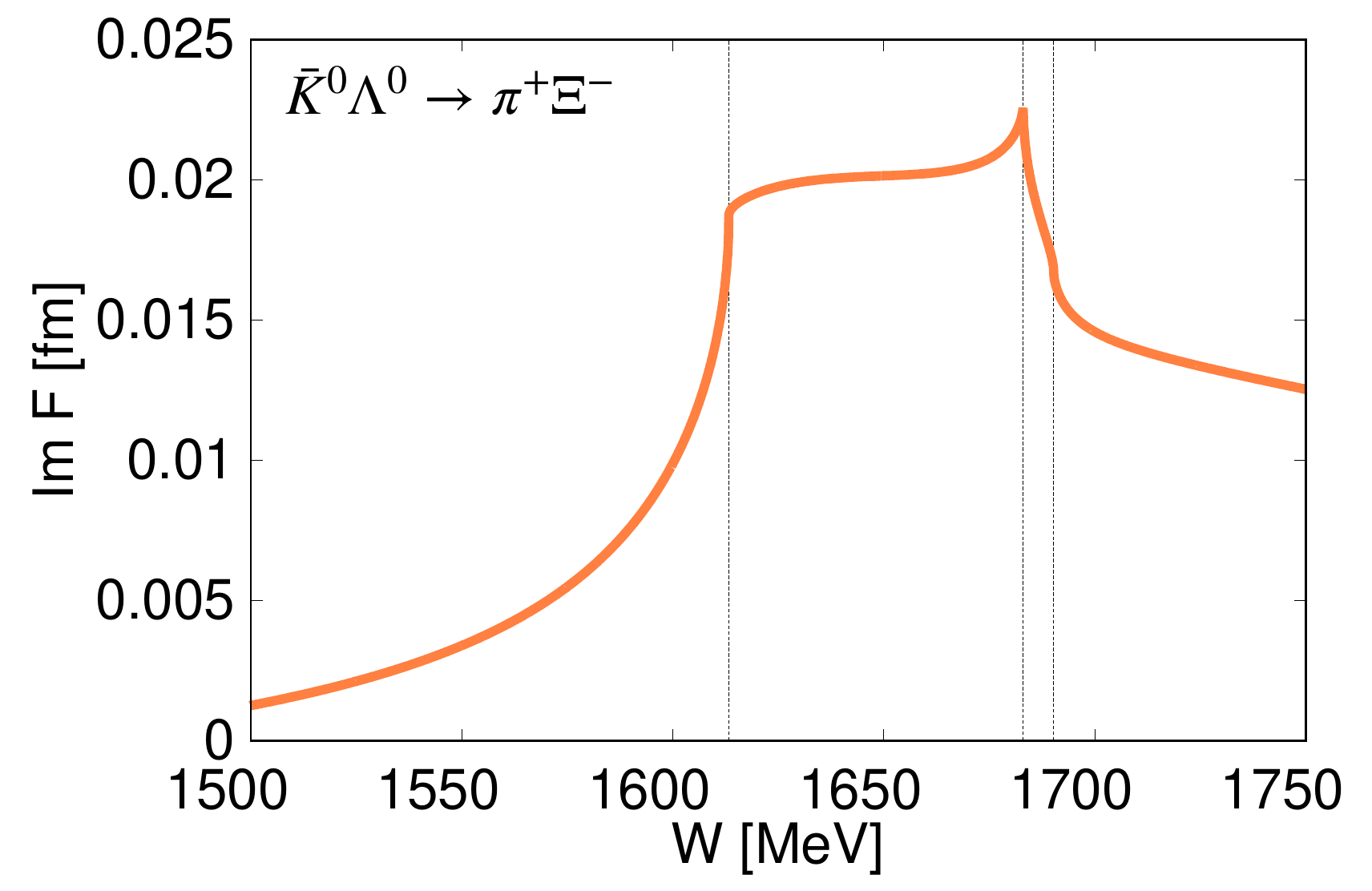}
\includegraphics[width=8cm]{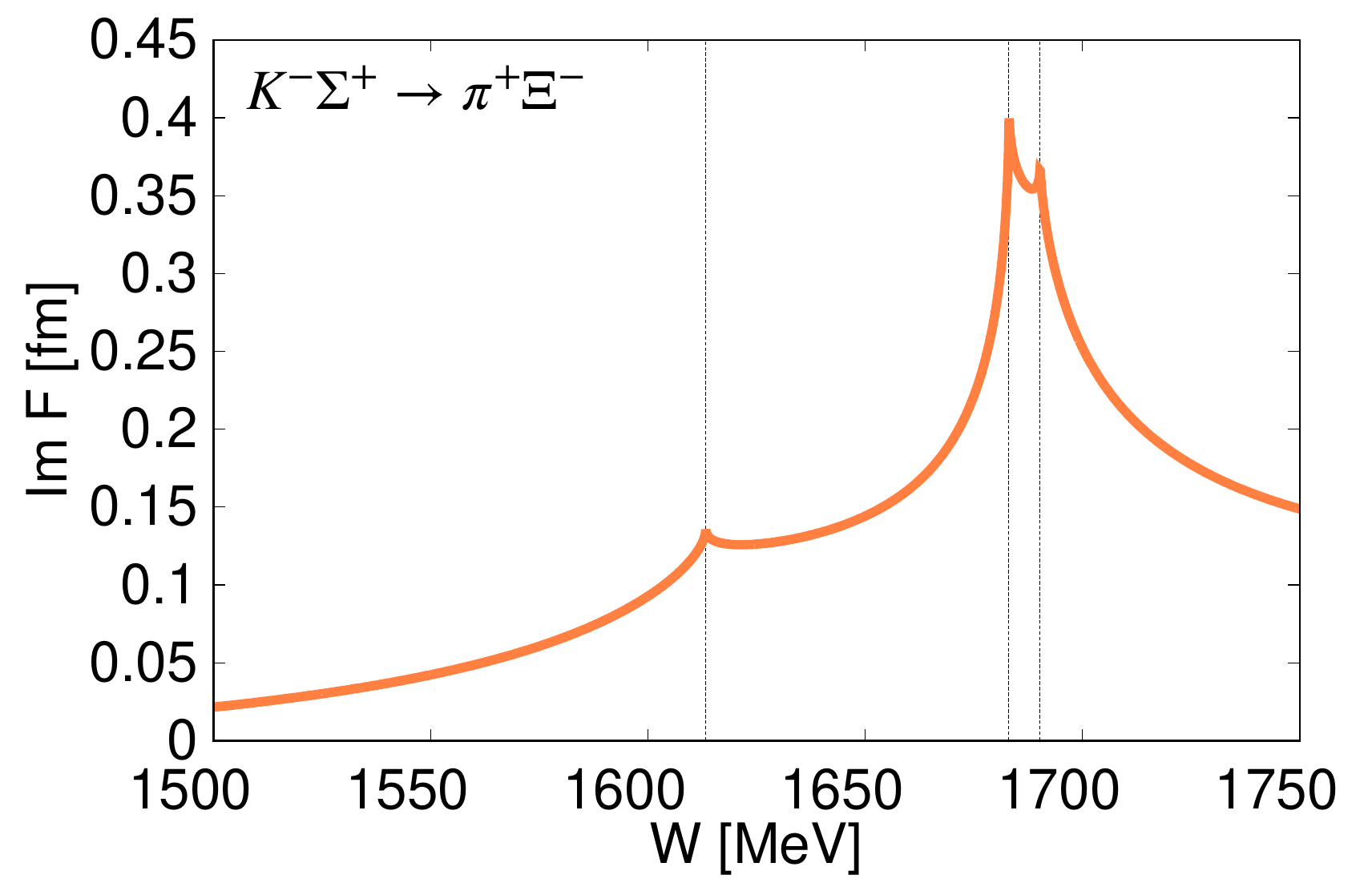}
\caption{
Transition scattering amplitudes for $\bar{K}^{0}\Lambda \to \pi^{+}\Xi^{-}$ (upper panel) and $K^{-}\Sigma^{+} \to \pi^{+}\Xi^{-}$ (lower panel) in Model 2. }
\label{fig:amplitude}
\end{figure}

From the expression of the transition amplitude $\mathcal{M}_{j}$ in Eq.~\eqref{eq:Minv}, it can be seen that the two-body scattering amplitudes $T_{ij}$ are summed over with the weights of the initial states $h_i$. Therefore, not only the diagonal components but also the off-diagonal components of the scattering amplitudes contribute to the invariant mass distribution.\footnote{In fact, in the $\Xi_{c}\to \pi\pi\Xi$ decay, the diagonal component does not contribute because of $h_{\pi^{+}\Xi^{-}}=0$.} Figure~\ref{fig:amplitude} shows the imaginary parts of the transition amplitudes for $\bar{K}^{0}\Lambda \to \pi^{+}\Xi^{-}$ and $K^{-}\Sigma^{+} \to \pi^{+}\Xi^{-}$. Unlike the diagonal components, a strong $\bar{K}\Sigma$ cusp structure is particularly visible in the $K^{-}\Sigma^{+} \to \pi^{+}\Xi^{-}$ channel. Since this $K^{-}\Sigma^{+}$ channel carries the largest weight $h_{K^{-}\Sigma^{+}} = 1$, it is expected to contribute significantly to the transition amplitude $\mathcal{M}_{j}$. Thus, the pronounced $\bar{K}\Sigma$ cusp structure observed in the invariant mass distribution can be interpreted as a consequence of the enhancement of the off-diagonal component due to the specific features of the decay process.

\section{Summary}
\label{sec:summary}

In this study, we have investigated the origin of various resonance poles in coupled-channel scattering and discussed their impact on the observable spectrum, taking $\Xi(1620)$ and $\Xi(1690)$ appearing in meson-baryon scattering amplitudes as examples. We first construct a simplified two-channel model to examine the effect of the decay channel on the pole trajectory, in which a bound state turns into a resonance state via a virtual state. As a result, we find that the pole originating from the bound state does not directly evolve into a resonance pole above the threshold, but rather undergoes an interchange with a pole originated from the virtual state.

Next, we have constructed a chiral unitary model to describe meson-baryon scattering with strangeness $S=-2$, and investigated the relationship between the $\Xi(1620)$ and $\Xi(1690)$ states described by different models by performing the interpolation between them. As a result, we find that even poles located in nearby energy regions and expected to share a common origin are not necessarily connected across models due to the structure of the Riemann sheets. Furthermore, we have demonstrated that the isospin symmetry breaking causes threshold energy differences between isospin partners, and when resonance poles appear near the isospin partner thresholds, the order of the model interpolation and the isospin breaking interpolation may not commute, leading to a transition between different poles.

Finally, we have calculated the $\pi^{+}\Xi^{-}$ invariant mass distribution in the $\Xi_{c} \to \pi\pi\Xi$ decay and discussed the impact of the poles on the observable spectrum along the real axis. Although many poles are found in the scattering amplitude, a peak structure appears in the spectrum only when a narrow-width pole exists on the physically relevant Riemann sheet. In contrast, poles with small imaginary parts located on other Riemann sheets do not lead to an enhancement of the scattering length in the threshold channel, and no peak appears in the spectrum. On the other hand, since off-diagonal transition amplitudes contribute to the invariant mass distribution, we have demonstrated that a strong cusp structure can arise even if the diagonal two-body scattering amplitude does not show the prominent cusp structure with moderate magnitude of the scattering length.

As a future perspective, it is desirable to examine the transition from a bound state to a resonance, investigated here within the chiral unitary model, in a more general framework of coupled-channel scattering. For the analysis of the $\Xi_{c} \to \pi\pi\Xi$ decay, it is also necessary to develop a more refined model, for example by including higher-order terms in chiral perturbation theory or incorporating the contribution from the $p$-wave resonance $\Xi(1530)$. The results of the present study are expected to provide a basis for such future developments.

\begin{acknowledgments}
This work has been supported in part by the Grants-in-Aid for Scientific Research from JSPS (Grants
No.~JP23H05439 and 
No.~JP22K03637), 
    and by JST SPRING, Grant Number JPMJSP2156, by JST, the establishment of university fellowships towards the creation of science technology innovation, Grant No. JPMJFS2139.
\end{acknowledgments}

\appendix
\section{Pole trajectory with the variation of coupling strength $\alpha$}
\label{app:poletrajectory}

In this appendix, we show the pole trajectories of the two-channel model in Sec.~\ref{sec:twochannel} with the variation of the coupling strength $\alpha$. We use the same setup with Sec.~\ref{sec:twochannel} with an attractive interaction strength $\alpha = 4$, a transition coupling strength $\beta=0$, and a subtraction constant $a_2 = -3$, which gives a bound state pole at 1495 MeV on the [tt] and [bt] sheets (pole 1) and a virtual state pole at 1436 MeV on the [tb] and [bb] sheets (pole 2). 

\begin{figure*}[tbp]
\centering
\includegraphics[width=8cm]{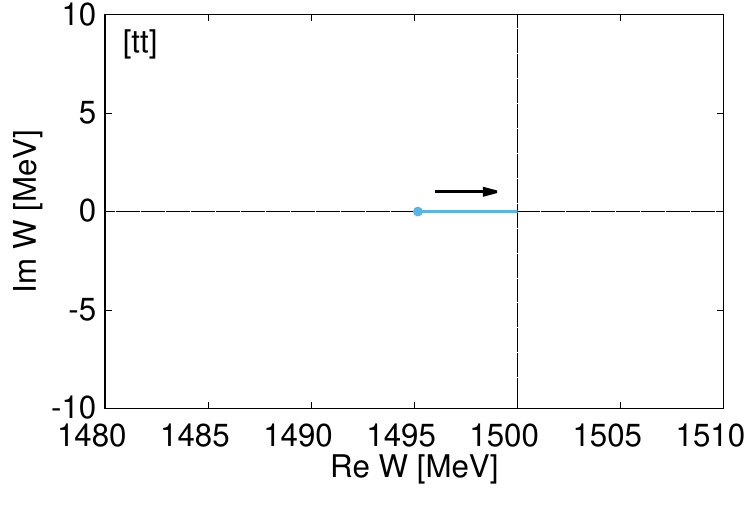}
\includegraphics[width=8cm]{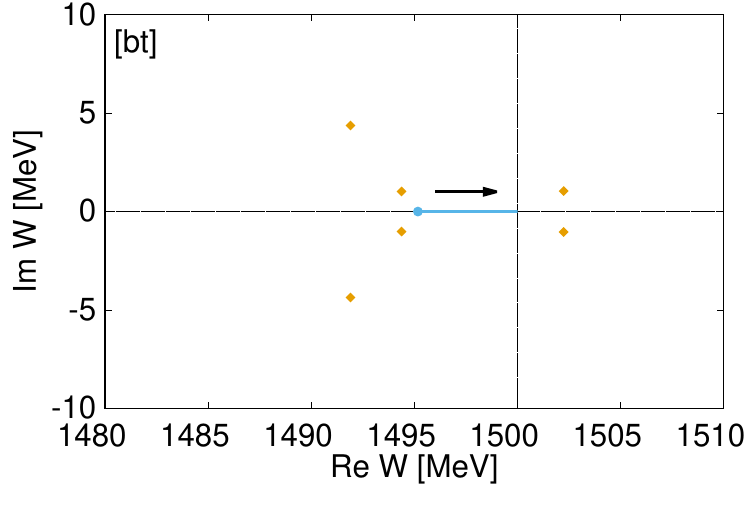}
\includegraphics[width=8cm]{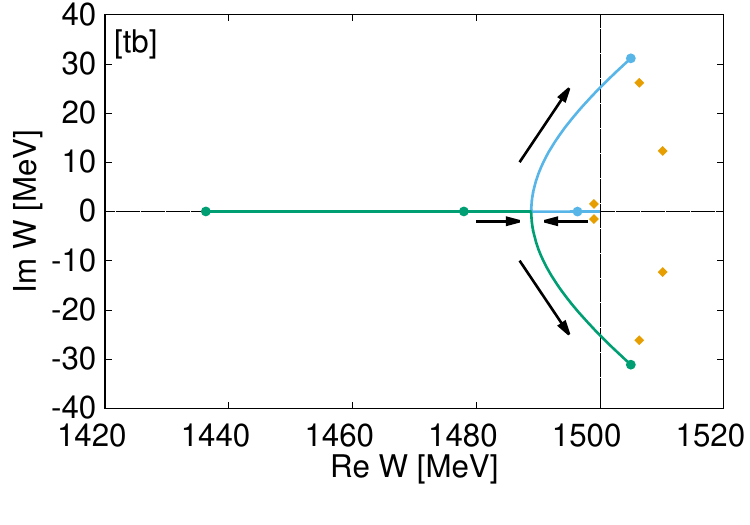}
\includegraphics[width=8cm]{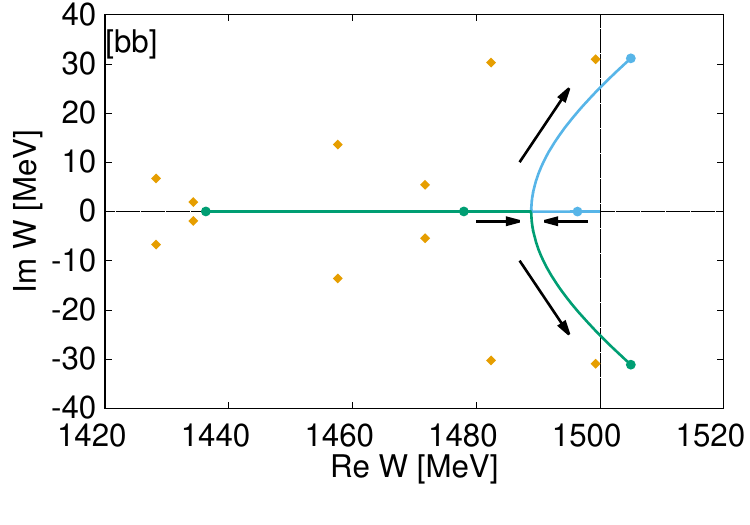}
\caption{Pole trajectories in the [tt] sheet (left top), [bt] sheet (right top), [tb] sheet (left bottom), [bb] sheet (right bottom), with the variation of the coupling strength $\alpha$ for $\beta=0$ and $a_{2}=-3$. The arrows indicate the direction of the pole movements  as $\alpha$ is decreased. Poles at $\alpha=4.0$, $3.4$, and $3.0$ are marked by the circles. Squares represent the pole positions with $\beta=0.4$ and $\beta=0.8$ (finite channel coupling).}
\label{fig:pole_beta0_alpha_E}
\end{figure*}

\begin{figure*}[tbp]
\centering
\includegraphics[width=8cm]{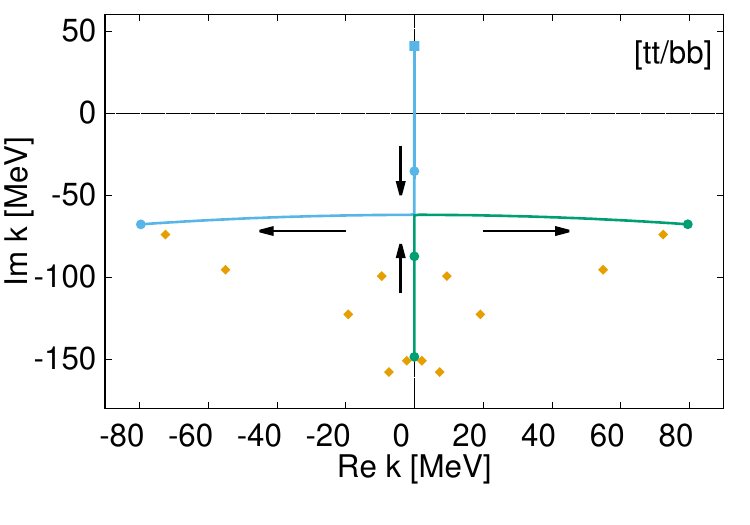}
\includegraphics[width=8cm]{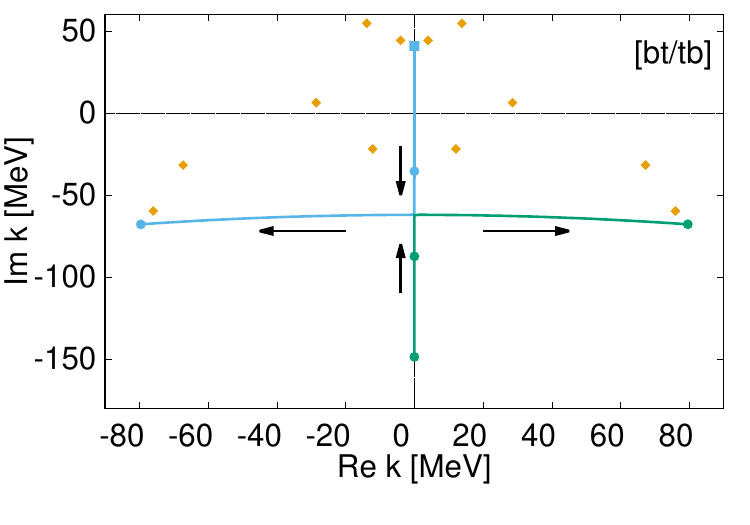}
\caption{Same with Fig.~\ref{fig:pole_beta0_alpha_E} but in the complex momentum [tt/bb] plane (left) and [bt/tb] plane (right).}
\label{fig:pole_beta0_alpha_k}
\end{figure*}

By decreasing the coupling constant $\alpha$ from $4.0$, both the bound state pole 1 and the virtual state pole 2 moves toward the threshold of channel 2. When $\alpha \sim 3.7$, pole 1 changes from a bound state to a virtual state, and at $\alpha \sim 3.35$, it collides with pole 2. A pair of the resonance and anti-resonance is observed at $\alpha = 3.0$. The corresponding pole trajectories are shown in Fig.~\ref{fig:pole_beta0_alpha_E} in the complex energy plane and Fig.~\ref{fig:pole_beta0_alpha_k} in the complex momentum plane. Representative pole positions for $\alpha = 4.0$ and $3.4$ (both poles are virtual states), and $3.0$ (resonance and anti-resonance) are summarized in Table~\ref{tab:pole_beta0_alpha}. Although there are some quantitative differences, the pole trajectories are qualitatively the same as those shown in Figs.~\ref{fig:pole_beta0_a2_E} and \ref{fig:pole_beta0_a2_k}.

\begin{table}[btp]
   \caption{Pole positions in the two-channel model with $\beta=0$ and $a_{2}=-3$. The bound state ($B$) appears in the [tt] and [bt] sheets, while the virtual state ($V$), resonance ($R$), and anti-resonance ($\bar{R}$) appear in the [tb] and [bb] sheets. }
   \begin{ruledtabular}
   \begin{tabular}{llll}
   $\alpha$ [dimensionless] & $4.0$      & $3.4$      & $3.0$ \\
   \hline
   Pole 1 [MeV]             & 1495 ($B$) & 1496 ($V$) & $1505+31i$ ($\bar{R}$) \\
   Pole 2 [MeV]             & 1436 ($V$) & 1477 ($V$) & $1505-31i$ ($R$)
   \end{tabular}
   \end{ruledtabular}
   \label{tab:pole_beta0_alpha}
\end{table}

To examine the effects of channel coupling, the pole positions for $\alpha = 4.0$, $3.4$, and $3.0$ with $\beta = 0.4$ and $0.8$ are indicated by squares in Figs.~\ref{fig:pole_beta0_alpha_E} and \ref{fig:pole_beta0_alpha_k}. The results are again qualitatively similar to those shown in Figs.~\ref{fig:pole_beta0_a2_E} and \ref{fig:pole_beta0_a2_k}. However, in the case of $\alpha = 3.4$ and $\beta = 0.8$, the poles evolved from pole~1 appears on the [bt] sheet in Fig.~\ref{fig:pole_beta0_alpha_E}, whereas it appeared on the [tb] sheet in Fig.~\ref{fig:pole_beta0_a2_E}. This can be understood by the poles in the complex momentum plane (right panel of Fig.~\ref{fig:pole_beta0_alpha_k}). When the virtual state originally located on the imaginary axis in the momentum plane is shifted into the complex plane due to channel coupling, the imaginary part decreases the pole reaches the upper half of the momentum plane at $\beta = 0.8$, and thus appears on the [bt] sheet.

\begin{figure*}[tbp]
\centering
\includegraphics[width=8cm]{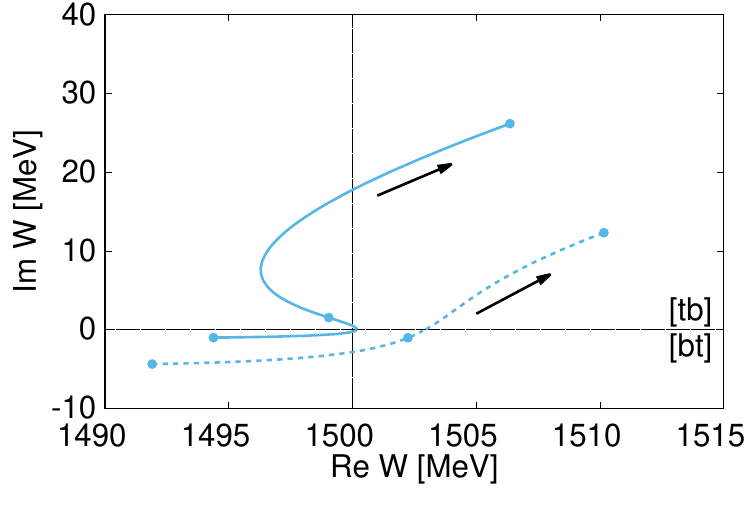}
\includegraphics[width=8cm]{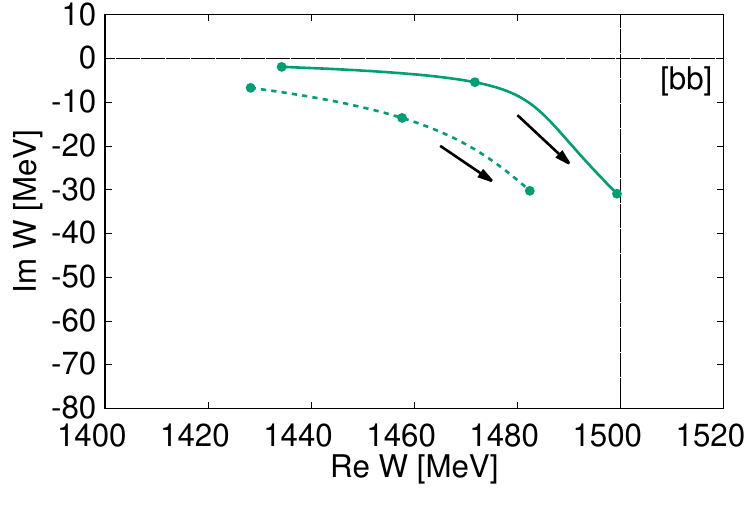}
\caption{Pole trajectories in the complex energy plane with the variation of the coupling strength $\alpha$ with $\beta=0.4$ (solid line) and $\beta=0.8$ (dashed line) for $a_{2}=-3.00$. The arrows indicate the direction of the pole movements as $\alpha$ is decreased. The left panel shows pole 1 on the [tb] sheet in the upper half plane and [bt] sheet in the lower half plane, while the right panel shows pole 2 on the [bb] sheet. The trajectories of the conjugate poles are omitted for simplicity. Poles at $\alpha=4.0$, $3.4$, and $3.0$ are marked by the circles.}
\label{fig:pole_beta_alpha_E}
\end{figure*}

\begin{figure}[tbp]
\centering
\includegraphics[width=8cm]{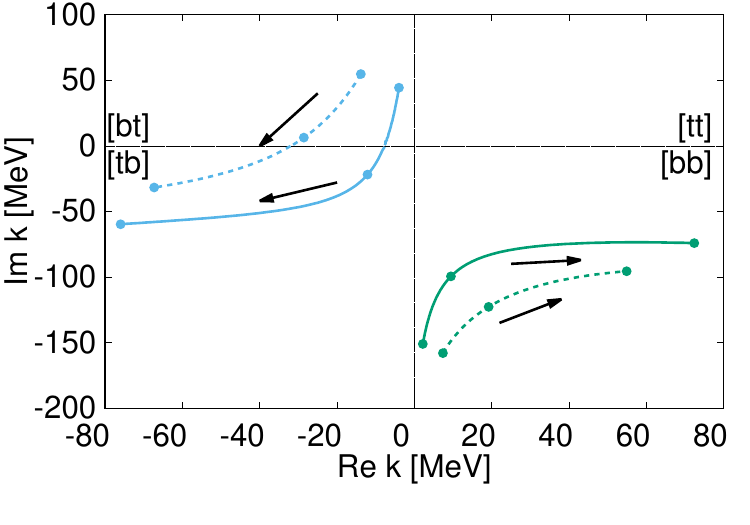}
\caption{Same with Fig.~\ref{fig:pole_beta_alpha_E} but in the complex momentum plane. In this figure, the [tt/bb] plane is plotted in the Re $k>0$ region, whereas the [bt/tb] plane is plotted in the Re $k<0$ region.}
\label{fig:pole_beta_alpha_k}
\end{figure}

Finally, we show the plot of the pole trajectories with the variation of $\alpha$ with fixed finite $\beta=0.4$ and $0.8$, in the complex energy plane (Fig.~\ref{fig:pole_beta_alpha_E}) and in the complex momentum plane (Fig.~\ref{fig:pole_beta_alpha_k}). In the case of $\beta = 0.4$, the quasibound state pole on the [bt] sheet moves into the [tb] sheet slightly above the threshold, loops around the threshold toward lower energies, and then turns to increase its real part, eventually turns into a shadow pole above the threshold. When $\beta$ is increased to 0.8, the overall behavior remains similar, but the looping around the threshold no longer appears. On the other hand, the quasivirtual state on the [bb] sheet moves toward lower energies as the interaction strength is reduced and eventually becomes a resonance pole above the threshold. In this way, the pole trajectories induced by varying the coupling strength also show an interchange between pole 1 and pole 2. This demonstrates that such a pole interchange is not a special feature of the subtraction constant variation discussed in the main text, but also occurs when the coupling strength is varied.

\begin{thebibliography}{40}%
\makeatletter
\providecommand \@ifxundefined [1]{%
 \@ifx{#1\undefined}
}%
\providecommand \@ifnum [1]{%
 \ifnum #1\expandafter \@firstoftwo
 \else \expandafter \@secondoftwo
 \fi
}%
\providecommand \@ifx [1]{%
 \ifx #1\expandafter \@firstoftwo
 \else \expandafter \@secondoftwo
 \fi
}%
\providecommand \natexlab [1]{#1}%
\providecommand \enquote  [1]{``#1''}%
\providecommand \bibnamefont  [1]{#1}%
\providecommand \bibfnamefont [1]{#1}%
\providecommand \citenamefont [1]{#1}%
\providecommand \href@noop [0]{\@secondoftwo}%
\providecommand \href [0]{\begingroup \@sanitize@url \@href}%
\providecommand \@href[1]{\@@startlink{#1}\@@href}%
\providecommand \@@href[1]{\endgroup#1\@@endlink}%
\providecommand \@sanitize@url [0]{\catcode `\\12\catcode `\$12\catcode
  `\&12\catcode `\#12\catcode `\^12\catcode `\_12\catcode `\%12\relax}%
\providecommand \@@startlink[1]{}%
\providecommand \@@endlink[0]{}%
\providecommand \url  [0]{\begingroup\@sanitize@url \@url }%
\providecommand \@url [1]{\endgroup\@href {#1}{\urlprefix }}%
\providecommand \urlprefix  [0]{URL }%
\providecommand \Eprint [0]{\href }%
\providecommand \doibase [0]{https://doi.org/}%
\providecommand \selectlanguage [0]{\@gobble}%
\providecommand \bibinfo  [0]{\@secondoftwo}%
\providecommand \bibfield  [0]{\@secondoftwo}%
\providecommand \translation [1]{[#1]}%
\providecommand \BibitemOpen [0]{}%
\providecommand \bibitemStop [0]{}%
\providecommand \bibitemNoStop [0]{.\EOS\space}%
\providecommand \EOS [0]{\spacefactor3000\relax}%
\providecommand \BibitemShut  [1]{\csname bibitem#1\endcsname}%
\let\auto@bib@innerbib\@empty
\bibitem [{\citenamefont {Navas}\ \emph {et~al.}(2024)\citenamefont {Navas}
  \emph {et~al.}}]{ParticleDataGroup:2024cfk}%
  \BibitemOpen
  \bibfield  {author} {\bibinfo {author} {\bibfnamefont {S.}~\bibnamefont
  {Navas}} \emph {et~al.} (\bibinfo {collaboration} {Particle Data Group}),\
  }\bibfield  {title} {\bibinfo {title} {{Review of particle physics}},\ }\href
  {https://doi.org/10.1103/PhysRevD.110.030001} {\bibfield  {journal} {\bibinfo
   {journal} {Phys. Rev. D}\ }\textbf {\bibinfo {volume} {110}},\ \bibinfo
  {pages} {030001} (\bibinfo {year} {2024})}\BibitemShut {NoStop}%
\bibitem [{\citenamefont {Hosaka}\ \emph {et~al.}(2016)\citenamefont {Hosaka},
  \citenamefont {Iijima}, \citenamefont {Miyabayashi}, \citenamefont {Sakai},\
  and\ \citenamefont {Yasui}}]{Hosaka:2016pey}%
  \BibitemOpen
  \bibfield  {author} {\bibinfo {author} {\bibfnamefont {A.}~\bibnamefont
  {Hosaka}}, \bibinfo {author} {\bibfnamefont {T.}~\bibnamefont {Iijima}},
  \bibinfo {author} {\bibfnamefont {K.}~\bibnamefont {Miyabayashi}}, \bibinfo
  {author} {\bibfnamefont {Y.}~\bibnamefont {Sakai}},\ and\ \bibinfo {author}
  {\bibfnamefont {S.}~\bibnamefont {Yasui}},\ }\bibfield  {title} {\bibinfo
  {title} {{Exotic hadrons with heavy flavors: X, Y, Z, and related states}},\
  }\href {https://doi.org/10.1093/ptep/ptw045} {\bibfield  {journal} {\bibinfo
  {journal} {PTEP}\ }\textbf {\bibinfo {volume} {2016}},\ \bibinfo {pages}
  {062C01} (\bibinfo {year} {2016})},\ \Eprint
  {https://arxiv.org/abs/1603.09229} {arXiv:1603.09229 [hep-ph]} \BibitemShut
  {NoStop}%
\bibitem [{\citenamefont {Brambilla}\ \emph {et~al.}(2020)\citenamefont
  {Brambilla}, \citenamefont {Eidelman}, \citenamefont {Hanhart}, \citenamefont
  {Nefediev}, \citenamefont {Shen}, \citenamefont {Thomas}, \citenamefont
  {Vairo},\ and\ \citenamefont {Yuan}}]{Brambilla:2019esw}%
  \BibitemOpen
  \bibfield  {author} {\bibinfo {author} {\bibfnamefont {N.}~\bibnamefont
  {Brambilla}}, \bibinfo {author} {\bibfnamefont {S.}~\bibnamefont {Eidelman}},
  \bibinfo {author} {\bibfnamefont {C.}~\bibnamefont {Hanhart}}, \bibinfo
  {author} {\bibfnamefont {A.}~\bibnamefont {Nefediev}}, \bibinfo {author}
  {\bibfnamefont {C.-P.}\ \bibnamefont {Shen}}, \bibinfo {author}
  {\bibfnamefont {C.~E.}\ \bibnamefont {Thomas}}, \bibinfo {author}
  {\bibfnamefont {A.}~\bibnamefont {Vairo}},\ and\ \bibinfo {author}
  {\bibfnamefont {C.-Z.}\ \bibnamefont {Yuan}},\ }\bibfield  {title} {\bibinfo
  {title} {{The $XYZ$ states: experimental and theoretical status and
  perspectives}},\ }\href {https://doi.org/10.1016/j.physrep.2020.05.001}
  {\bibfield  {journal} {\bibinfo  {journal} {Phys. Rept.}\ }\textbf {\bibinfo
  {volume} {873}},\ \bibinfo {pages} {1} (\bibinfo {year} {2020})},\ \Eprint
  {https://arxiv.org/abs/1907.07583} {arXiv:1907.07583 [hep-ex]} \BibitemShut
  {NoStop}%
\bibitem [{\citenamefont {Guo}\ \emph {et~al.}(2018)\citenamefont {Guo},
  \citenamefont {Hanhart}, \citenamefont {Mei{\ss}ner}, \citenamefont {Wang},
  \citenamefont {Zhao},\ and\ \citenamefont {Zou}}]{Guo:2017jvc}%
  \BibitemOpen
  \bibfield  {author} {\bibinfo {author} {\bibfnamefont {F.-K.}\ \bibnamefont
  {Guo}}, \bibinfo {author} {\bibfnamefont {C.}~\bibnamefont {Hanhart}},
  \bibinfo {author} {\bibfnamefont {U.-G.}\ \bibnamefont {Mei{\ss}ner}},
  \bibinfo {author} {\bibfnamefont {Q.}~\bibnamefont {Wang}}, \bibinfo {author}
  {\bibfnamefont {Q.}~\bibnamefont {Zhao}},\ and\ \bibinfo {author}
  {\bibfnamefont {B.-S.}\ \bibnamefont {Zou}},\ }\bibfield  {title} {\bibinfo
  {title} {{Hadronic molecules}},\ }\href
  {https://doi.org/10.1103/RevModPhys.90.015004} {\bibfield  {journal}
  {\bibinfo  {journal} {Rev. Mod. Phys.}\ }\textbf {\bibinfo {volume} {90}},\
  \bibinfo {pages} {015004} (\bibinfo {year} {2018})},\ \bibinfo {note}
  {[Erratum: Rev.Mod.Phys. 94, 029901 (2022)]},\ \Eprint
  {https://arxiv.org/abs/1705.00141} {arXiv:1705.00141 [hep-ph]} \BibitemShut
  {NoStop}%
\bibitem [{\citenamefont {Hyodo}\ and\ \citenamefont
  {Niiyama}(2021)}]{Hyodo:2020czb}%
  \BibitemOpen
  \bibfield  {author} {\bibinfo {author} {\bibfnamefont {T.}~\bibnamefont
  {Hyodo}}\ and\ \bibinfo {author} {\bibfnamefont {M.}~\bibnamefont
  {Niiyama}},\ }\bibfield  {title} {\bibinfo {title} {{QCD and the strange
  baryon spectrum}},\ }\href {https://doi.org/10.1016/j.ppnp.2021.103868}
  {\bibfield  {journal} {\bibinfo  {journal} {Prog. Part. Nucl. Phys.}\
  }\textbf {\bibinfo {volume} {120}},\ \bibinfo {pages} {103868} (\bibinfo
  {year} {2021})},\ \Eprint {https://arxiv.org/abs/2010.07592}
  {arXiv:2010.07592 [hep-ph]} \BibitemShut {NoStop}%
\bibitem [{\citenamefont {Taylor}(1972)}]{Taylor}%
  \BibitemOpen
  \bibfield  {author} {\bibinfo {author} {\bibfnamefont {J.~R.}\ \bibnamefont
  {Taylor}},\ }\href@noop {} {\emph {\bibinfo {title} {Scattering Theory: The
  Quantum Theory on Nonrelativistic Collisions}}}\ (\bibinfo  {publisher}
  {Wiley},\ \bibinfo {address} {New York},\ \bibinfo {year} {1972})\BibitemShut
  {NoStop}%
\bibitem [{\citenamefont {Rakityansky}(2022)}]{Rakityansky}%
  \BibitemOpen
  \bibfield  {author} {\bibinfo {author} {\bibfnamefont {S.~A.}\ \bibnamefont
  {Rakityansky}},\ }\href@noop {} {\emph {\bibinfo {title} {Jost Functions in
  Quantum Mechanics}}}\ (\bibinfo  {publisher} {Springer},\ \bibinfo {year}
  {2022})\BibitemShut {NoStop}%
\bibitem [{\citenamefont {Mai}\ \emph {et~al.}(2023)\citenamefont {Mai},
  \citenamefont {Mei\ss{}ner},\ and\ \citenamefont {Urbach}}]{Mai:2022eur}%
  \BibitemOpen
  \bibfield  {author} {\bibinfo {author} {\bibfnamefont {M.}~\bibnamefont
  {Mai}}, \bibinfo {author} {\bibfnamefont {U.-G.}\ \bibnamefont
  {Mei\ss{}ner}},\ and\ \bibinfo {author} {\bibfnamefont {C.}~\bibnamefont
  {Urbach}},\ }\bibfield  {title} {\bibinfo {title} {{Towards a theory of
  hadron resonances}},\ }\href@noop {} {\bibfield  {journal} {\bibinfo
  {journal} {Phys. Rept.}\ }\textbf {\bibinfo {volume} {1001}},\ \bibinfo
  {pages} {1} (\bibinfo {year} {2023})},\ \Eprint
  {https://arxiv.org/abs/2206.01477} {arXiv:2206.01477 [hep-ph]} \BibitemShut
  {NoStop}%
\bibitem [{\citenamefont {Mai}(2025)}]{Mai:2025wjb}%
  \BibitemOpen
  \bibfield  {author} {\bibinfo {author} {\bibfnamefont {M.}~\bibnamefont
  {Mai}},\ }\bibfield  {title} {\bibinfo {title} {{Theory of resonances}},\
  }\href@noop {} {\  (\bibinfo {year} {2025})},\ \Eprint
  {https://arxiv.org/abs/2502.02654} {arXiv:2502.02654 [hep-ph]} \BibitemShut
  {NoStop}%
\bibitem [{\citenamefont {Nishibuchi}\ and\ \citenamefont
  {Hyodo}(2024)}]{Nishibuchi:2023acl}%
  \BibitemOpen
  \bibfield  {author} {\bibinfo {author} {\bibfnamefont {T.}~\bibnamefont
  {Nishibuchi}}\ and\ \bibinfo {author} {\bibfnamefont {T.}~\bibnamefont
  {Hyodo}},\ }\bibfield  {title} {\bibinfo {title} {{Analysis of the
  $\Xi(1620)$ resonance and $\bar{K}\Lambda$ scattering length with a chiral
  unitary approach}},\ }\href@noop {} {\bibfield  {journal} {\bibinfo
  {journal} {Phys. Rev. C}\ }\textbf {\bibinfo {volume} {109}},\ \bibinfo
  {pages} {015203} (\bibinfo {year} {2024})},\ \Eprint
  {https://arxiv.org/abs/2305.10753} {arXiv:2305.10753 [hep-ph]} \BibitemShut
  {NoStop}%
\bibitem [{\citenamefont {Hyodo}(2014)}]{Hyodo:2014bda}%
  \BibitemOpen
  \bibfield  {author} {\bibinfo {author} {\bibfnamefont {T.}~\bibnamefont
  {Hyodo}},\ }\bibfield  {title} {\bibinfo {title} {{Hadron mass scaling near
  the s-wave threshold}},\ }\href@noop {} {\bibfield  {journal} {\bibinfo
  {journal} {Phys. Rev. C}\ }\textbf {\bibinfo {volume} {90}},\ \bibinfo
  {pages} {055208} (\bibinfo {year} {2014})},\ \Eprint
  {https://arxiv.org/abs/1407.2372} {arXiv:1407.2372 [hep-ph]} \BibitemShut
  {NoStop}%
\bibitem [{\citenamefont {Hanhart}\ \emph {et~al.}(2014)\citenamefont
  {Hanhart}, \citenamefont {Pelaez},\ and\ \citenamefont
  {Rios}}]{Hanhart:2014ssa}%
  \BibitemOpen
  \bibfield  {author} {\bibinfo {author} {\bibfnamefont {C.}~\bibnamefont
  {Hanhart}}, \bibinfo {author} {\bibfnamefont {J.}~\bibnamefont {Pelaez}},\
  and\ \bibinfo {author} {\bibfnamefont {G.}~\bibnamefont {Rios}},\ }\bibfield
  {title} {\bibinfo {title} {{Remarks on pole trajectories for resonances}},\
  }\href@noop {} {\bibfield  {journal} {\bibinfo  {journal} {Phys. Lett. B}\
  }\textbf {\bibinfo {volume} {739}},\ \bibinfo {pages} {375} (\bibinfo {year}
  {2014})},\ \Eprint {https://arxiv.org/abs/1407.7452} {arXiv:1407.7452
  [hep-ph]} \BibitemShut {NoStop}%
\bibitem [{\citenamefont {Sumihama}\ \emph {et~al.}(2019)\citenamefont
  {Sumihama} \emph {et~al.}}]{Belle:2018lws}%
  \BibitemOpen
  \bibfield  {author} {\bibinfo {author} {\bibfnamefont {M.}~\bibnamefont
  {Sumihama}} \emph {et~al.} (\bibinfo {collaboration} {Belle}),\ }\bibfield
  {title} {\bibinfo {title} {{Observation of $\Xi(1620)^0$ and evidence for
  $\Xi(1690)^0$ in $\Xi_c^+ \rightarrow \Xi^-\pi^+\pi^+$ decays}},\ }\href@noop
  {} {\bibfield  {journal} {\bibinfo  {journal} {Phys. Rev. Lett.}\ }\textbf
  {\bibinfo {volume} {122}},\ \bibinfo {pages} {072501} (\bibinfo {year}
  {2019})},\ \Eprint {https://arxiv.org/abs/1810.06181} {arXiv:1810.06181
  [hep-ex]} \BibitemShut {NoStop}%
\bibitem [{\citenamefont {Acharya}\ \emph {et~al.}(2021)\citenamefont {Acharya}
  \emph {et~al.}}]{ALICE:2020wvi}%
  \BibitemOpen
  \bibfield  {author} {\bibinfo {author} {\bibfnamefont {S.}~\bibnamefont
  {Acharya}} \emph {et~al.} (\bibinfo {collaboration} {ALICE}),\ }\bibfield
  {title} {\bibinfo {title} {{$\Lambda\rm{K}$ femtoscopy in Pb-Pb collisions at
  $\sqrt{s_{\rm{NN}}}$ = 2.76 TeV}},\ }\href@noop {} {\bibfield  {journal}
  {\bibinfo  {journal} {Phys. Rev. C}\ }\textbf {\bibinfo {volume} {103}},\
  \bibinfo {pages} {055201} (\bibinfo {year} {2021})},\ \Eprint
  {https://arxiv.org/abs/2005.11124} {arXiv:2005.11124 [nucl-ex]} \BibitemShut
  {NoStop}%
\bibitem [{\citenamefont {Acharya}\ \emph {et~al.}(2023)\citenamefont {Acharya}
  \emph {et~al.}}]{ALICE:2023wjz}%
  \BibitemOpen
  \bibfield  {author} {\bibinfo {author} {\bibfnamefont {S.}~\bibnamefont
  {Acharya}} \emph {et~al.} (\bibinfo {collaboration} {ALICE}),\ }\bibfield
  {title} {\bibinfo {title} {{Accessing the strong interaction between
  \ensuremath{\Lambda} baryons and charged kaons with the femtoscopy technique
  at the LHC}},\ }\href@noop {} {\bibfield  {journal} {\bibinfo  {journal}
  {Phys. Lett. B}\ }\textbf {\bibinfo {volume} {845}},\ \bibinfo {pages}
  {138145} (\bibinfo {year} {2023})},\ \Eprint
  {https://arxiv.org/abs/2305.19093} {arXiv:2305.19093 [nucl-ex]} \BibitemShut
  {NoStop}%
\bibitem [{\citenamefont {Kaiser}\ \emph {et~al.}(1995)\citenamefont {Kaiser},
  \citenamefont {Siegel},\ and\ \citenamefont {Weise}}]{Kaiser:1995eg}%
  \BibitemOpen
  \bibfield  {author} {\bibinfo {author} {\bibfnamefont {N.}~\bibnamefont
  {Kaiser}}, \bibinfo {author} {\bibfnamefont {P.~B.}\ \bibnamefont {Siegel}},\
  and\ \bibinfo {author} {\bibfnamefont {W.}~\bibnamefont {Weise}},\ }\bibfield
   {title} {\bibinfo {title} {Chiral dynamics and the low-energy kaon--nucleon
  interaction},\ }\href@noop {} {\bibfield  {journal} {\bibinfo  {journal}
  {Nucl. Phys. A}\ }\textbf {\bibinfo {volume} {594}},\ \bibinfo {pages} {325}
  (\bibinfo {year} {1995})},\ \Eprint {https://arxiv.org/abs/nucl-th/9505043}
  {nucl-th/9505043} \BibitemShut {NoStop}%
\bibitem [{\citenamefont {Oset}\ and\ \citenamefont
  {Ramos}(1998)}]{Oset:1997it}%
  \BibitemOpen
  \bibfield  {author} {\bibinfo {author} {\bibfnamefont {E.}~\bibnamefont
  {Oset}}\ and\ \bibinfo {author} {\bibfnamefont {A.}~\bibnamefont {Ramos}},\
  }\bibfield  {title} {\bibinfo {title} {{Nonperturbative chiral approach to s
  wave anti-K N interactions}},\ }\href
  {https://doi.org/10.1016/S0375-9474(98)00170-5} {\bibfield  {journal}
  {\bibinfo  {journal} {Nucl. Phys.}\ }\textbf {\bibinfo {volume} {A635}},\
  \bibinfo {pages} {99} (\bibinfo {year} {1998})},\ \Eprint
  {https://arxiv.org/abs/nucl-th/9711022} {arXiv:nucl-th/9711022 [nucl-th]}
  \BibitemShut {NoStop}%
\bibitem [{\citenamefont {Oller}\ and\ \citenamefont
  {Meissner}(2001)}]{Oller:2000fj}%
  \BibitemOpen
  \bibfield  {author} {\bibinfo {author} {\bibfnamefont {J.~A.}\ \bibnamefont
  {Oller}}\ and\ \bibinfo {author} {\bibfnamefont {U.~G.}\ \bibnamefont
  {Meissner}},\ }\bibfield  {title} {\bibinfo {title} {Chiral dynamics in the
  presence of bound states: kaon--nucleon interactions revisited},\ }\href@noop
  {} {\bibfield  {journal} {\bibinfo  {journal} {Phys. Lett. B}\ }\textbf
  {\bibinfo {volume} {500}},\ \bibinfo {pages} {263} (\bibinfo {year}
  {2001})},\ \Eprint {https://arxiv.org/abs/hep-ph/0011146} {hep-ph/0011146}
  \BibitemShut {NoStop}%
\bibitem [{\citenamefont {Hyodo}\ and\ \citenamefont
  {Jido}(2012)}]{Hyodo:2011ur}%
  \BibitemOpen
  \bibfield  {author} {\bibinfo {author} {\bibfnamefont {T.}~\bibnamefont
  {Hyodo}}\ and\ \bibinfo {author} {\bibfnamefont {D.}~\bibnamefont {Jido}},\
  }\bibfield  {title} {\bibinfo {title} {{The nature of the $\Lambda(1405)$
  resonance in chiral dynamics}},\ }\href@noop {} {\bibfield  {journal}
  {\bibinfo  {journal} {Prog. Part. Nucl. Phys.}\ }\textbf {\bibinfo {volume}
  {67}},\ \bibinfo {pages} {55} (\bibinfo {year} {2012})},\ \Eprint
  {https://arxiv.org/abs/1104.4474} {arXiv:1104.4474 [nucl-th]} \BibitemShut
  {NoStop}%
\bibitem [{\citenamefont {Ramos}\ \emph {et~al.}(2002)\citenamefont {Ramos},
  \citenamefont {Oset},\ and\ \citenamefont {Bennhold}}]{Ramos:2002xh}%
  \BibitemOpen
  \bibfield  {author} {\bibinfo {author} {\bibfnamefont {A.}~\bibnamefont
  {Ramos}}, \bibinfo {author} {\bibfnamefont {E.}~\bibnamefont {Oset}},\ and\
  \bibinfo {author} {\bibfnamefont {C.}~\bibnamefont {Bennhold}},\ }\bibfield
  {title} {\bibinfo {title} {On the spin, parity and nature of the xi(1620)
  resonance},\ }\href
  {http://www.slac.stanford.edu/spires/find/hep/www?eprint=NUCL-TH 0204044}
  {\bibfield  {journal} {\bibinfo  {journal} {Phys. Rev. Lett.}\ }\textbf
  {\bibinfo {volume} {89}},\ \bibinfo {pages} {252001} (\bibinfo {year}
  {2002})},\ \Eprint {https://arxiv.org/abs/nucl-th/0204044} {nucl-th/0204044}
  \BibitemShut {NoStop}%
\bibitem [{\citenamefont {Garcia-Recio}\ \emph {et~al.}(2004)\citenamefont
  {Garcia-Recio}, \citenamefont {Lutz},\ and\ \citenamefont
  {Nieves}}]{Garcia-Recio:2003ejq}%
  \BibitemOpen
  \bibfield  {author} {\bibinfo {author} {\bibfnamefont {C.}~\bibnamefont
  {Garcia-Recio}}, \bibinfo {author} {\bibfnamefont {M.~F.~M.}\ \bibnamefont
  {Lutz}},\ and\ \bibinfo {author} {\bibfnamefont {J.}~\bibnamefont {Nieves}},\
  }\bibfield  {title} {\bibinfo {title} {{Quark mass dependence of s wave
  baryon resonances}},\ }\href@noop {} {\bibfield  {journal} {\bibinfo
  {journal} {Phys. Lett. B}\ }\textbf {\bibinfo {volume} {582}},\ \bibinfo
  {pages} {49} (\bibinfo {year} {2004})},\ \Eprint
  {https://arxiv.org/abs/nucl-th/0305100} {arXiv:nucl-th/0305100} \BibitemShut
  {NoStop}%
\bibitem [{\citenamefont {Sekihara}(2015)}]{Sekihara:2015qqa}%
  \BibitemOpen
  \bibfield  {author} {\bibinfo {author} {\bibfnamefont {T.}~\bibnamefont
  {Sekihara}},\ }\bibfield  {title} {\bibinfo {title} {{$\Xi (1690)$ as a
  $\bar{K} \Sigma$ molecular state}},\ }\href@noop {} {\bibfield  {journal}
  {\bibinfo  {journal} {PTEP}\ }\textbf {\bibinfo {volume} {2015}},\ \bibinfo
  {pages} {091D01} (\bibinfo {year} {2015})},\ \Eprint
  {https://arxiv.org/abs/1505.02849} {arXiv:1505.02849 [hep-ph]} \BibitemShut
  {NoStop}%
\bibitem [{\citenamefont {Khemchandani}\ \emph {et~al.}(2018)\citenamefont
  {Khemchandani}, \citenamefont {Martinez~Torres}, \citenamefont {Hosaka},
  \citenamefont {Nagahiro}, \citenamefont {Navarra},\ and\ \citenamefont
  {Nielsen}}]{Khemchandani:2016ftn}%
  \BibitemOpen
  \bibfield  {author} {\bibinfo {author} {\bibfnamefont {K.~P.}\ \bibnamefont
  {Khemchandani}}, \bibinfo {author} {\bibfnamefont {A.}~\bibnamefont
  {Martinez~Torres}}, \bibinfo {author} {\bibfnamefont {A.}~\bibnamefont
  {Hosaka}}, \bibinfo {author} {\bibfnamefont {H.}~\bibnamefont {Nagahiro}},
  \bibinfo {author} {\bibfnamefont {F.~S.}\ \bibnamefont {Navarra}},\ and\
  \bibinfo {author} {\bibfnamefont {M.}~\bibnamefont {Nielsen}},\ }\bibfield
  {title} {\bibinfo {title} {{Why $\Xi(1690)$ and $\Xi(2120)$ are so
  narrow?}},\ }\href@noop {} {\bibfield  {journal} {\bibinfo  {journal} {Phys.
  Rev.}\ }\textbf {\bibinfo {volume} {D97}},\ \bibinfo {pages} {034005}
  (\bibinfo {year} {2018})},\ \Eprint {https://arxiv.org/abs/1608.07086}
  {arXiv:1608.07086 [nucl-th]} \BibitemShut {NoStop}%
\bibitem [{\citenamefont {Feijoo}\ \emph {et~al.}(2023)\citenamefont {Feijoo},
  \citenamefont {Valcarce~Cadenas},\ and\ \citenamefont
  {Magas}}]{Feijoo:2023wua}%
  \BibitemOpen
  \bibfield  {author} {\bibinfo {author} {\bibfnamefont {A.}~\bibnamefont
  {Feijoo}}, \bibinfo {author} {\bibfnamefont {V.}~\bibnamefont
  {Valcarce~Cadenas}},\ and\ \bibinfo {author} {\bibfnamefont {V.~K.}\
  \bibnamefont {Magas}},\ }\bibfield  {title} {\bibinfo {title} {{The
  $\Xi(1620)$ and $\Xi(1690)$ molecular states from $S=-2$ meson-baryon
  interaction up to next-to-leading order}},\ }\href@noop {} {\bibfield
  {journal} {\bibinfo  {journal} {Phys. Lett. B}\ }\textbf {\bibinfo {volume}
  {841}},\ \bibinfo {pages} {137927} (\bibinfo {year} {2023})},\ \Eprint
  {https://arxiv.org/abs/2303.01323} {arXiv:2303.01323 [hep-ph]} \BibitemShut
  {NoStop}%
\bibitem [{\citenamefont {Li}\ \emph {et~al.}(2023)\citenamefont {Li},
  \citenamefont {Zhang}, \citenamefont {Liang},\ and\ \citenamefont
  {Oset}}]{Li:2023olv}%
  \BibitemOpen
  \bibfield  {author} {\bibinfo {author} {\bibfnamefont {H.-P.}\ \bibnamefont
  {Li}}, \bibinfo {author} {\bibfnamefont {G.-J.}\ \bibnamefont {Zhang}},
  \bibinfo {author} {\bibfnamefont {W.-H.}\ \bibnamefont {Liang}},\ and\
  \bibinfo {author} {\bibfnamefont {E.}~\bibnamefont {Oset}},\ }\bibfield
  {title} {\bibinfo {title} {{Theoretical interpretation of the $\Xi (1620)$
  and $\Xi (1690)$ resonances seen in $\Xi _c^+ \rightarrow \Xi ^- \pi ^+ \pi
  ^+$ decay}},\ }\href@noop {} {\bibfield  {journal} {\bibinfo  {journal} {Eur.
  Phys. J. C}\ }\textbf {\bibinfo {volume} {83}},\ \bibinfo {pages} {954}
  (\bibinfo {year} {2023})},\ \Eprint {https://arxiv.org/abs/2308.11879}
  {arXiv:2308.11879 [hep-ph]} \BibitemShut {NoStop}%
\bibitem [{\citenamefont {Sarti}\ \emph {et~al.}(2024)\citenamefont {Sarti},
  \citenamefont {Feijoo}, \citenamefont {Vida\~na}, \citenamefont {Ramos},
  \citenamefont {Giacosa}, \citenamefont {Hyodo},\ and\ \citenamefont
  {Kamiya}}]{Sarti:2023wlg}%
  \BibitemOpen
  \bibfield  {author} {\bibinfo {author} {\bibfnamefont {V.~M.}\ \bibnamefont
  {Sarti}}, \bibinfo {author} {\bibfnamefont {A.}~\bibnamefont {Feijoo}},
  \bibinfo {author} {\bibfnamefont {I.}~\bibnamefont {Vida\~na}}, \bibinfo
  {author} {\bibfnamefont {A.}~\bibnamefont {Ramos}}, \bibinfo {author}
  {\bibfnamefont {F.}~\bibnamefont {Giacosa}}, \bibinfo {author} {\bibfnamefont
  {T.}~\bibnamefont {Hyodo}},\ and\ \bibinfo {author} {\bibfnamefont
  {Y.}~\bibnamefont {Kamiya}},\ }\bibfield  {title} {\bibinfo {title}
  {{Constraining the low-energy S=-2 meson-baryon interaction with two-particle
  correlations}},\ }\href@noop {} {\bibfield  {journal} {\bibinfo  {journal}
  {Phys. Rev. D}\ }\textbf {\bibinfo {volume} {110}},\ \bibinfo {pages}
  {L011505} (\bibinfo {year} {2024})},\ \Eprint
  {https://arxiv.org/abs/2309.08756} {arXiv:2309.08756 [hep-ph]} \BibitemShut
  {NoStop}%
\bibitem [{\citenamefont {Feijoo}\ \emph {et~al.}(2025)\citenamefont {Feijoo},
  \citenamefont {Sarti}, \citenamefont {Nieves}, \citenamefont {Ramos},\ and\
  \citenamefont {Vida\~na}}]{Feijoo:2024qqg}%
  \BibitemOpen
  \bibfield  {author} {\bibinfo {author} {\bibfnamefont {A.}~\bibnamefont
  {Feijoo}}, \bibinfo {author} {\bibfnamefont {V.~M.}\ \bibnamefont {Sarti}},
  \bibinfo {author} {\bibfnamefont {J.}~\bibnamefont {Nieves}}, \bibinfo
  {author} {\bibfnamefont {A.}~\bibnamefont {Ramos}},\ and\ \bibinfo {author}
  {\bibfnamefont {I.}~\bibnamefont {Vida\~na}},\ }\bibfield  {title} {\bibinfo
  {title} {{Bridging correlation and spectroscopy measurements to access the
  hadron interaction behind molecular states: The case of the
  \ensuremath{\Xi}(1620) and \ensuremath{\Xi}(1690) in the
  K-\ensuremath{\Lambda} system}},\ }\href@noop {} {\bibfield  {journal}
  {\bibinfo  {journal} {Phys. Rev. D}\ }\textbf {\bibinfo {volume} {111}},\
  \bibinfo {pages} {014022} (\bibinfo {year} {2025})},\ \Eprint
  {https://arxiv.org/abs/2411.10245} {arXiv:2411.10245 [hep-ph]} \BibitemShut
  {NoStop}%
\bibitem [{\citenamefont {Miyahara}\ \emph {et~al.}(2015)\citenamefont
  {Miyahara}, \citenamefont {Hyodo},\ and\ \citenamefont
  {Oset}}]{Miyahara:2015cja}%
  \BibitemOpen
  \bibfield  {author} {\bibinfo {author} {\bibfnamefont {K.}~\bibnamefont
  {Miyahara}}, \bibinfo {author} {\bibfnamefont {T.}~\bibnamefont {Hyodo}},\
  and\ \bibinfo {author} {\bibfnamefont {E.}~\bibnamefont {Oset}},\ }\bibfield
  {title} {\bibinfo {title} {{Weak decay of $\Lambda_{c}^+$ for the study of
  $\Lambda$(1405) and $\Lambda$(1670)}},\ }\href@noop {} {\bibfield  {journal}
  {\bibinfo  {journal} {Phys. Rev. C}\ }\textbf {\bibinfo {volume} {92}},\
  \bibinfo {pages} {055204} (\bibinfo {year} {2015})},\ \Eprint
  {https://arxiv.org/abs/1508.04882} {arXiv:1508.04882 [nucl-th]} \BibitemShut
  {NoStop}%
\bibitem [{\citenamefont {Miyahara}\ \emph {et~al.}(2017)\citenamefont
  {Miyahara}, \citenamefont {Hyodo}, \citenamefont {Oka}, \citenamefont
  {Nieves},\ and\ \citenamefont {Oset}}]{Miyahara:2016yyh}%
  \BibitemOpen
  \bibfield  {author} {\bibinfo {author} {\bibfnamefont {K.}~\bibnamefont
  {Miyahara}}, \bibinfo {author} {\bibfnamefont {T.}~\bibnamefont {Hyodo}},
  \bibinfo {author} {\bibfnamefont {M.}~\bibnamefont {Oka}}, \bibinfo {author}
  {\bibfnamefont {J.}~\bibnamefont {Nieves}},\ and\ \bibinfo {author}
  {\bibfnamefont {E.}~\bibnamefont {Oset}},\ }\bibfield  {title} {\bibinfo
  {title} {{Theoretical study of the $\Xi(1620)$ and $\Xi(1690)$ resonances in
  $\Xi_c \to \pi^+ MB$ decays}},\ }\href@noop {} {\bibfield  {journal}
  {\bibinfo  {journal} {Phys. Rev. C}\ }\textbf {\bibinfo {volume} {95}},\
  \bibinfo {pages} {035212} (\bibinfo {year} {2017})},\ \Eprint
  {https://arxiv.org/abs/1609.00895} {arXiv:1609.00895 [nucl-th]} \BibitemShut
  {NoStop}%
\bibitem [{\citenamefont {Oset}\ \emph {et~al.}(2016)\citenamefont {Oset} \emph
  {et~al.}}]{Oset:2016lyh}%
  \BibitemOpen
  \bibfield  {author} {\bibinfo {author} {\bibfnamefont {E.}~\bibnamefont
  {Oset}} \emph {et~al.},\ }\bibfield  {title} {\bibinfo {title} {{Weak decays
  of heavy hadrons into dynamically generated resonances}},\ }\href
  {https://doi.org/10.1142/S0218301316300010} {\bibfield  {journal} {\bibinfo
  {journal} {Int. J. Mod. Phys.}\ }\textbf {\bibinfo {volume} {E25}},\ \bibinfo
  {pages} {1630001} (\bibinfo {year} {2016})},\ \Eprint
  {https://arxiv.org/abs/1601.03972} {arXiv:1601.03972 [hep-ph]} \BibitemShut
  {NoStop}%
\bibitem [{\citenamefont {Hyodo}\ \emph {et~al.}(2008)\citenamefont {Hyodo},
  \citenamefont {Jido},\ and\ \citenamefont {Hosaka}}]{Hyodo:2008xr}%
  \BibitemOpen
  \bibfield  {author} {\bibinfo {author} {\bibfnamefont {T.}~\bibnamefont
  {Hyodo}}, \bibinfo {author} {\bibfnamefont {D.}~\bibnamefont {Jido}},\ and\
  \bibinfo {author} {\bibfnamefont {A.}~\bibnamefont {Hosaka}},\ }\bibfield
  {title} {\bibinfo {title} {{Origin of resonances in the chiral unitary
  approach}},\ }\href@noop {} {\bibfield  {journal} {\bibinfo  {journal} {Phys.
  Rev. C}\ }\textbf {\bibinfo {volume} {78}},\ \bibinfo {pages} {025203}
  (\bibinfo {year} {2008})},\ \Eprint {https://arxiv.org/abs/0803.2550}
  {arXiv:0803.2550 [nucl-th]} \BibitemShut {NoStop}%
\bibitem [{\citenamefont {Ikeda}\ \emph {et~al.}(2011)\citenamefont {Ikeda},
  \citenamefont {Hyodo}, \citenamefont {Jido}, \citenamefont {Kamano},
  \citenamefont {Sato},\ and\ \citenamefont {Yazaki}}]{Ikeda:2011dx}%
  \BibitemOpen
  \bibfield  {author} {\bibinfo {author} {\bibfnamefont {Y.}~\bibnamefont
  {Ikeda}}, \bibinfo {author} {\bibfnamefont {T.}~\bibnamefont {Hyodo}},
  \bibinfo {author} {\bibfnamefont {D.}~\bibnamefont {Jido}}, \bibinfo {author}
  {\bibfnamefont {H.}~\bibnamefont {Kamano}}, \bibinfo {author} {\bibfnamefont
  {T.}~\bibnamefont {Sato}},\ and\ \bibinfo {author} {\bibfnamefont
  {K.}~\bibnamefont {Yazaki}},\ }\bibfield  {title} {\bibinfo {title}
  {{Structure of $\Lambda(1405)$ and threshold behavior of $\pi\Sigma$
  scattering}},\ }\href@noop {} {\bibfield  {journal} {\bibinfo  {journal}
  {Prog. Theor. Phys.}\ }\textbf {\bibinfo {volume} {125}},\ \bibinfo {pages}
  {1205} (\bibinfo {year} {2011})},\ \Eprint {https://arxiv.org/abs/1101.5190}
  {arXiv:1101.5190 [nucl-th]} \BibitemShut {NoStop}%
\bibitem [{\citenamefont {Hyodo}(2013)}]{Hyodo:2013iga}%
  \BibitemOpen
  \bibfield  {author} {\bibinfo {author} {\bibfnamefont {T.}~\bibnamefont
  {Hyodo}},\ }\bibfield  {title} {\bibinfo {title} {{Structure of
  Near-Threshold s-Wave Resonances}},\ }\href@noop {} {\bibfield  {journal}
  {\bibinfo  {journal} {Phys. Rev. Lett.}\ }\textbf {\bibinfo {volume} {111}},\
  \bibinfo {pages} {132002} (\bibinfo {year} {2013})},\ \Eprint
  {https://arxiv.org/abs/1305.1999} {arXiv:1305.1999 [hep-ph]} \BibitemShut
  {NoStop}%
\bibitem [{\citenamefont {W.D.Heiss}(1999)}]{PRE.61.929}%
  \BibitemOpen
  \bibfield  {author} {\bibinfo {author} {\bibnamefont {W.D.Heiss}},\
  }\bibfield  {title} {\bibinfo {title} {{Repulsion of resonance states and
  exceptional points}},\ }\href {https://doi.org/10.1103/PhysRevE.61.929}
  {\bibfield  {journal} {\bibinfo  {journal} {Phys. Rev. E}\ }\textbf {\bibinfo
  {volume} {61}},\ \bibinfo {pages} {929} (\bibinfo {year} {1999})},\ \Eprint
  {https://arxiv.org/abs/9909047} {arXiv:9909047 [quant-ph]} \BibitemShut
  {NoStop}%
\bibitem [{\citenamefont {Heiss}(2012)}]{Heiss:2012dx}%
  \BibitemOpen
  \bibfield  {author} {\bibinfo {author} {\bibfnamefont {W.~D.}\ \bibnamefont
  {Heiss}},\ }\bibfield  {title} {\bibinfo {title} {{The physics of exceptional
  points}},\ }\href@noop {} {\bibfield  {journal} {\bibinfo  {journal} {J.
  Phys. A}\ }\textbf {\bibinfo {volume} {45}},\ \bibinfo {pages} {444016}
  (\bibinfo {year} {2012})},\ \Eprint {https://arxiv.org/abs/1210.7536}
  {arXiv:1210.7536 [quant-ph]} \BibitemShut {NoStop}%
\bibitem [{\citenamefont {Moiseyev}(2011)}]{Moiseyev}%
  \BibitemOpen
  \bibfield  {author} {\bibinfo {author} {\bibfnamefont {N.}~\bibnamefont
  {Moiseyev}},\ }\href@noop {} {\emph {\bibinfo {title} {Non-Hermitian Quantum
  Mechanics}}}\ (\bibinfo  {publisher} {Cambridge University Press},\ \bibinfo
  {address} {Cambridge},\ \bibinfo {year} {2011})\BibitemShut {NoStop}%
\bibitem [{\citenamefont {Eden}\ and\ \citenamefont
  {Taylor}(1964)}]{Eden:1964zz}%
  \BibitemOpen
  \bibfield  {author} {\bibinfo {author} {\bibfnamefont {R.}~\bibnamefont
  {Eden}}\ and\ \bibinfo {author} {\bibfnamefont {J.}~\bibnamefont {Taylor}},\
  }\bibfield  {title} {\bibinfo {title} {{Poles and Shadow Poles in the
  Many-Channel S Matrix}},\ }\href {https://doi.org/10.1103/PhysRev.133.B1575}
  {\bibfield  {journal} {\bibinfo  {journal} {Phys. Rev.}\ }\textbf {\bibinfo
  {volume} {133}},\ \bibinfo {pages} {B1575} (\bibinfo {year}
  {1964})}\BibitemShut {NoStop}%
\bibitem [{\citenamefont {Kamiya}\ and\ \citenamefont
  {Hyodo}(2018)}]{Kamiya:2017pcq}%
  \BibitemOpen
  \bibfield  {author} {\bibinfo {author} {\bibfnamefont {Y.}~\bibnamefont
  {Kamiya}}\ and\ \bibinfo {author} {\bibfnamefont {T.}~\bibnamefont {Hyodo}},\
  }\bibfield  {title} {\bibinfo {title} {{Structure of hadron resonances with a
  nearby zero of the amplitude}},\ }\href@noop {} {\bibfield  {journal}
  {\bibinfo  {journal} {Phys. Rev.}\ }\textbf {\bibinfo {volume} {D97}},\
  \bibinfo {pages} {054019} (\bibinfo {year} {2018})},\ \Eprint
  {https://arxiv.org/abs/1711.04558} {arXiv:1711.04558 [hep-ph]} \BibitemShut
  {NoStop}%
\bibitem [{\citenamefont {Baru}\ \emph {et~al.}(2005)\citenamefont {Baru},
  \citenamefont {Haidenbauer}, \citenamefont {Hanhart}, \citenamefont
  {Kudryavtsev},\ and\ \citenamefont {Meissner}}]{Baru:2004xg}%
  \BibitemOpen
  \bibfield  {author} {\bibinfo {author} {\bibfnamefont {V.}~\bibnamefont
  {Baru}}, \bibinfo {author} {\bibfnamefont {J.}~\bibnamefont {Haidenbauer}},
  \bibinfo {author} {\bibfnamefont {C.}~\bibnamefont {Hanhart}}, \bibinfo
  {author} {\bibfnamefont {A.~E.}\ \bibnamefont {Kudryavtsev}},\ and\ \bibinfo
  {author} {\bibfnamefont {U.-G.}\ \bibnamefont {Meissner}},\ }\bibfield
  {title} {\bibinfo {title} {{Flatt\'e-like distributions and the $a_0(980) /
  f_0(980)$ mesons}},\ }\href@noop {} {\bibfield  {journal} {\bibinfo
  {journal} {Eur. Phys. J. A}\ }\textbf {\bibinfo {volume} {23}},\ \bibinfo
  {pages} {523} (\bibinfo {year} {2005})},\ \Eprint
  {https://arxiv.org/abs/nucl-th/0410099} {arXiv:nucl-th/0410099} \BibitemShut
  {NoStop}%
\bibitem [{\citenamefont {Sone}\ and\ \citenamefont
  {Hyodo}(2024)}]{Sone:2024nfj}%
  \BibitemOpen
  \bibfield  {author} {\bibinfo {author} {\bibfnamefont {K.}~\bibnamefont
  {Sone}}\ and\ \bibinfo {author} {\bibfnamefont {T.}~\bibnamefont {Hyodo}},\
  }\bibfield  {title} {\bibinfo {title} {{General amplitude of near-threshold
  hadron scattering for exotic hadrons}},\ }\href@noop {} {\  (\bibinfo {year}
  {2024})},\ \Eprint {https://arxiv.org/abs/2405.08436} {arXiv:2405.08436
  [hep-ph]} \BibitemShut {NoStop}%
\end{thebibliography}
%

\end{document}